\documentclass[aps,prc,preprint,showpacs,preprintnumbers,nofootinbib,float,superscriptaddress,longbibliography]{revtex4-1}
\pdfoutput=1
\usepackage{graphicx, fancybox}
\usepackage{amsmath,amssymb}
\usepackage[colorlinks=true, pdfstartview=FitV, linkcolor=red,
		     citecolor=blue, urlcolor=blue]{hyperref}
\usepackage{color}
\usepackage{soul}
\usepackage{url}
\usepackage[normalem]{ulem}
\usepackage{nccmath}
\usepackage{amsthm}
\usepackage{mathrsfs}
\usepackage{graphicx}
\usepackage{caption}
\usepackage{subcaption}
\usepackage{float}

\begin{document}
   
\title{\bf Dynamical gauge fields and anomalous transport at strong coupling}
\author{A.D. Gallegos$^1$ and U. G\"ursoy}
\affiliation{Institute for Theoretical Physics, Utrecht University, Leuvenlaan 4, 3584 CE Utrecht, The Netherlands}
\date{\today}


\begin{abstract}

Anomalous transport coefficients are known to be universal in the absence of dynamical gauge fields. We calculate the corrections to these universal values due to dynamical gluon fields at strong coupling, at finite temperature and finite density, using the holographic duality. We show that the consistent chiral magnetic and chiral vortical currents receive no corrections, while we derive a semi-analytic formula for the chiral separation conductivity. We determine these corrections in the large color, large flavor limit, in terms of a series expansion in the anomalous dimension $\Delta$ of  the axial current  in terms of physical parameters $\Delta$, temperature, electric and chiral chemical potentials and the flavor to color ratio $\frac{N_f}{N_c}$. Our results are applicable to a generic class of chiral gauge theories that allow for a holographic description in the gravity approximation. We also determine the dynamical gluon corrections to the chiral vortical separation current in a particular example in the absence of external axial fields. 
\end{abstract}

\maketitle

\newpage

\section{Introduction and Summary}
\label{sec::intro}

In spite of a long history, dating back to \cite{Vilenkin:1980fu}, study of unusual transport phenomena induced by chiral anomalies flourished only recently with the hope of discovering such phenomena in real systems such as the quark gluon plasma produced in the heavy ion collisions at RHIC and LHC \cite{OriginalSon,OriginalKharzeev,Kharzeev:2007jp}\footnote{See the review \cite{LandsteinerReview} for a recent account.}. The canonical examples of anomalous transport are the chiral magnetic and vortical effects (CME and CVE), that refer to generation of a macroscopic electric current as a result of an axial anomaly in the presence of magnetic field $\mathbf{B}$ or vorticity $\boldsymbol{\omega}$ respectively. Whether evidence for the CME and CVE  can indeed be found in the heavy ion collisions is still an open issue \cite{Liao:2016diz, Kharzeev:2015znc, Adamczyk:2015eqo,Belmont:2014lta} but there exists strong evidence that anomalous transport finds experimental realization in Dirac and Weyl semimetals \cite{Li:2014bha,Zhang:2016ufu,Shekhar:2015rqa,Zhang:2015lya,Kim:2013dia,Cortijo:2016wnf}. More theoretically, such anomalous transport phenomena can also be confirmed by lattice QCD studies \cite{Yamamoto:2011gk,Braguta:2013loa}. 

One should distinguish between the two different types of anomalies that contribute to anomalous transport in a chiral gauge theory: i) anomalies caused by external fields, and ii) anomalies caused by dynamical gauge fields. Both appear in the conservation equation for an abelian axial current $J_A$ as follows, 
\begin{equation}\label{aneq}
\nabla_\mu  J^\mu_A  = \frac{\epsilon^{\mu \nu \rho \sigma}}{4} \left[ a_1 F^V_{\mu \nu} F^V_{\rho \sigma} + a_2 F^A_{\mu \nu} F^A_{\rho \sigma} + a_3 \text{Tr}\left(G_{\mu \nu} G_{\rho \sigma} \right)  + a_4 R^\alpha_{\hphantom{\alpha} \beta \mu \nu} R^\beta_{\hphantom{\beta} \alpha \rho \sigma}  \right]\, , 
\end{equation}
where $F^{V}=dV$ and $F^A=dA$ are the field strengths of external vector and axial gauge fields, $G$ is the field strength of the dynamical gauge fields in the theory, e.g. gluons, and $R^\alpha_{\hphantom{\alpha} \beta \mu \nu}$ is the Riemann tensor of the background geometry.  The anomaly coefficients $a_1$, $a_2$ and $a_4$ are examples of the first type whereas $a_3$ is of the second type. All are known to be one loop exact \cite{Adler:1969er}. 

The purpose of our paper is to explore the contribution of the second, {\em dynamical}, type of anomalies to anomalous transport in strongly coupled chiral gauge theories. We study the problem using the AdS/CFT correspondence \cite{Maldacena:1997re,Gubser:1998bc, Witten:1998qj}.   In particular we calculate, using AdS/CFT, the transport coefficients that characterize the anomalous transport properties of the system, including the contribution from the dynamical type anomalies. 

Anomaly induced axial and vector currents are given in the following Ohm form: 
\begin{equation}
\begin{split}
\mathbf{J}_{\text{\tiny V}} = \sigma_{\text{\tiny VV}} \mathbf{B} + \sigma_{\text{\tiny VA}} \mathbf{B}_5 + \sigma_{\text{\tiny$V\Omega$}} \boldsymbol{ \omega}, \\
\mathbf{J}_{\text{\tiny A}} = \sigma_{\text{\tiny AV}} \mathbf{B} + \sigma_{\text{\tiny AA}} \mathbf{B}_5 + \sigma_{\text{\tiny$A \Omega$}} \boldsymbol{ \omega}, 
\end{split}
\end{equation}
where $\mathbf{B}$,  $\mathbf{B}_5$ and $\mathbf{\omega}$ are the external vector and axial magnetic fields and vorticity respectively\footnote{Even though there are no {\em fundamental} axial magnetic fields in Nature, including them as sources is instrumental in calculation of the $\sigma_{AV}$ conductivity. Furthermore, they appear in the effective description of Weyl semimetals.}. Non-vanishing values of the conductivities $\{\sigma_{\text{\tiny VV}},\sigma_{\text{\tiny AV}},\sigma_{\text{ \tiny V$\Omega$}},\sigma_{\text{ \tiny A$\Omega$}} \}$ lead to the Chiral Magnetic Effect (CME), the Chiral Separation Effect (CSE), the Chiral Vortical Effect (CVE) and the Chiral Vortical Separation Effect (CVSE) respectively. In equilibrium these conductivities can be calculated from first principles via a Kubo type formula\cite{LandsteinerKubo}
%
\begin{equation}\label{KuboFormulas}
\begin{split}
  \sigma_{\text{\tiny MN}}  =   \lim_{k \rightarrow 0} \tilde \epsilon_{ijk}  \frac{i k^j}{2 k^2} \left. \langle J^i_{\text{\tiny M}} J^k_{\text{\tiny N}} \rangle  \right|_{\omega=0}, \quad \quad 
 \sigma_{\text{\tiny $M \Omega$ }} =  \lim_{k \rightarrow 0} \tilde \epsilon_{ijk} \frac{i k^j}{2 k^2} \left.  \langle J^i_{\text{\tiny M}} T^{tk} \rangle  \right|_{\omega=0},
\end{split}
\end{equation}
%
where the latin indices indicate spatial components, the indices \{M,N\}=\{A,V\} indicate the type. The two point functions are evaluated at exactly zero frequency. This condition allow us to equate $\sigma_{AV}$ and $\sigma_{VA}$, thus the vector one point function carries all the information regarding the CME, CSE and CVE conductivities. Alternatively one can obtain the conductivities from the linear response of the one-point functions to the magnetic like sources \cite{GursoyTarrio,Grozdanov}, which is the route we take in this work. In particular we read off the vector like conductivities from the linear response of the one-point function $\langle J_V \rangle$, 
\begin{equation}\label{coefs}
\langle  \delta  J^\nu_V \rangle = \sigma_{\text{\tiny VV}} B^\nu + \sigma_{\text{\tiny VA}} B^\nu_5 + \sigma_{\text{\tiny V$\Omega$}} \omega^\nu.   
\end{equation}


In the absence of dynamical contribution to the anomaly equation (\ref{aneq}), i.e. when $a_3=0$,  the anomalous transport coefficients in (\ref{coefs}) are universally determined by the values of $a_1$, $a_2$ and $a_4$ in a given QFT. For example in QCD coupled to external vector and axial gauge fields one finds for the {\em chiral magnetic}, {\em chiral separation} and  {\em chiral vortical} conductivities for the {\em covariant current}\footnote{The story of the {\em chiral vortical separation} conductivity is more involved. It involves a term quadratic in temperature, whose coefficient is only very recently understood to be protected by the {\em global gravitational anomaly} \cite{SonGolkar, Chowdhury:2016cmh, Golkar:2015oxw,Glorioso:2017lcn}.} : 
\begin{equation}
\label{CovCond} 
\sigma_{VV} = \frac{\mu_5}{2\pi^2}\, , \qquad \sigma_{AV} = \sigma_{VA} = \frac{\mu}{2\pi^2}\, ,\qquad \sigma_{V\Omega} = \frac{\mu\mu_5}{2\pi^2}\, , 
\end{equation}
where $\mu_5$ and $\mu$ are the axial and electric chemical potentials and we set the electric charge $e=1$. By ``universality'' we mean a) the same form (with only the coefficients vary depending on anomaly coeffcients) in any chiral gauge theory and b) non-renormalization of these values by radiative corrections. Universality is suggested by several field theoretic arguments, such as the energy balance \cite{Nielsen:1983rb}, Dirac index \cite{Metlitski:2005pr}, thermodynamic potential \cite{OriginalKharzeev} and derivative expansion of the effective action \cite{DHoker:1985ycn} that are all nicely summarized in \cite{OriginalKharzeev}. 

Same expressions follow from non-field theoretic approaches, such as the hydrodynamics \cite{Son:2009tf, Neiman:2010zi}, holography \cite{Erdmenger, Landsteiner1,Landsteiner2,Landsteiner3} and effective field theory \cite{Jensen:2012kj, Jensen:2013vta,Jensen:2013rga}. In particular, Son and Surowka  \cite{Son:2009tf} obtained the same results in the hydrodynamic regime with the additional assumption of a local entropy current with non-negative divergence.  Eventually the universality of the CME, CVE and CSE  coefficients in holography based on two-derivative gravity was established in \cite{GursoyTarrio} where they are linked to smooth near-horizon geometry of the corresponding black hole solutions\footnote{We will make use of this smoothness condition in this paper as well.}. This holographic demonstration was later extended to higher derivative gravity theories in \cite{Grozdanov}. 

In passing, we note that the actual values of anomalous conductivities depend on the choice of {\em covariant} current (BRST invariant) versus {\em consistent} current (satisfying the Wess-Zumino consistency conditions \cite{Wess:1971yu}). The two differ by the Chern-Simons current\footnote{See \cite{LandsteinerReview} for a careful recent presentation in the context of holography.}. We calculate the one-point function of the {\em consistent} current in (\ref{coefs}) in this paper, while most of the early literature  on the subject, which we review above, involved the {\em covariant} currents.
We present the {\em consistent} current version of the universal values\footnote{The CME conductivity should vanish in accordance to Bloch's theorem \cite{YamamotoBloch} for a system at equilibrium. The CVE conductivity, in fact, escapes this theorem because a rotating system cannot be in equilibrium in a relativistic theory. The apparent vanishing of the {\em consistent} CVE conductivity might just be a choice of fluid frame. To check this one can calculate the  energy current $T^{tk}$. This is beyond the scope of our paper. We thank Karl Landsteiner for pointing this out.} for the CME, CSE and CVE conductivities in (\ref{CovCond}) 
\begin{equation}
\label{ConsCond} 
\sigma_{VV} = 0\, , \qquad \sigma_{AV} = \sigma_{VA} = \frac{\mu}{2\pi^2}\, ,\qquad \sigma_{V\Omega} = 0\, , 
\end{equation}
which follow from (\ref{CovCond}) upon addition of the Chern-Simons current contribution. 

These values are modified when dynamical gauge fields are included, i.e. when $a_3$ in the anomaly equation (\ref{aneq}) is non-vanishing. Indeed it is known that once $a_3 \neq 0$ the anomalous conductivities will receive radiative corrections \cite{Gorbar:2011ya,Gorbar:2009bm,Gorbar:2010kc, Fukushima:2010zza, Gorbar:2013upa,Hou:2012xg,SonGolkar,renormalizationEffective}. This is also clear from the fact that the universal values (\ref{CovCond}) one finds in the absence of dynamical gluons disagree with lattice QCD calculations, which clearly include such contributions \cite{Yamamoto:2011gk,Braguta:2013loa}. Contribution of dynamical gauge fields to anomaly induced transport recently have been studied in thermal field theory \cite{Hou:2012xg, SonGolkar}, effective field theory \cite{renormalizationEffective}, and holography \cite{Aaron,LandsteinerStuckleberg}. In \cite{SonGolkar} a two loop correction to the CVSE conductivity was found while in \cite{renormalizationEffective} it was argued that all conductivities should receive corrections once the gluons are turned on. In \cite{Aaron}, the holographic dual of dynamical gauge field contribution is shown to be a bulk axion with a Stuckelberg coupling to the bulk gauge field corresponding to the axial current on the boundary. This idea was utilized in \cite{LandsteinerStuckleberg} to estimate such corrections to CME and CSE conductivities in the case of the conformal plasma. 

In general the hydrodynamic approach becomes ambiguous in the presence of dynamical gauge fields since it  is based on conservation of charge and the $a_3$ term in equation (\ref{aneq}) violates this inherently\footnote{Unlike the other terms in (\ref{aneq}) that can be turned off by turning off external sources.}. 

However, there is no obstruction in 't Hooft's large color limit---which is the limit of interest in holography. In this limit, the thermal average of the topological charge $\langle G \wedge G \rangle$ in the (deconfined) plasma phase vanish exponentially as $\exp(-N_c \times const.)$ as the only possible contribution from instantons are suppressed as such. This is established both in the lattice studies \cite{Vicari:2008jw} and in holography \cite{Gursoy:2008za}. It is important to note that, the exponential suppression of topological charge does not imply exponential suppression of the associated corrections to anomalous transport; akin to the famous result of Witten\cite{Witten:1978bc} and Veneziano \cite{Veneziano:1979ec} that the two-point function of topological charge is suppressed only by $1/N_c$,  rather than exponentially, giving an ${\cal O}(N_f/N_c)$ mass to $\eta'$, where $N_c$ and $N_f$ refer to the ranks of the gauge and flavor groups $SU(N_c)$ and $U(N_f)$. 

However, as noted in \cite{Aaron} that the dynamical corrections should, in fact, be suppressed as $N_f/N_c$ in the 't Hooft large color limit, just as the mass of the $\eta'$ meson. Thus we are forced to consider the Veneziano limit instead:
\begin{equation}
\label{Veneziano}
N_c \to \infty, \qquad N_f\to\infty, \qquad x \equiv \frac{N_f}{N_c} = const. 
\end{equation} 
In the holographic bulk picture, sending $N_f$ to infinity necessitates backreaction of the flavor branes to the background geometry, that complicates the gravitational system substantially \cite{Bigazzi:2005md,Casero:2006pt,Jarvinen:2011qe}. 

In this paper, assuming such a gravitational background at finite $x$, we calculate the chiral magnetic, separation and vortical conductivities, including  corrections from dynamical gauge fields, following the holographic prescription developed in \cite{Aaron,LandsteinerStuckleberg}.  In particular, we organize the holographic calculation in a series expansion in the anomalous dimension of the axial current $\Delta$ and obtain generic formulas---which can be applied to a generic background---for corrections to the consistent currents at first order in $\Delta$.  Our main results are that, at this order in $\Delta$ the CME and CVE conductivities remain the same  as in (\ref{ConsCond}) and (\ref{CovCond}) whereas the CSE conductivity is modified, given by equation (\ref{CSE}). 

The structure of the paper is as follows. In the next two sections we set the stage for the  holographic calculation of the anomaly induced conductivities in the presence of dynamical gauge fields. In particular, in section \ref{HolographicSetup} we explain how the bulk axion is related to the dynamical gluons in the dual field theory, and in section \ref{HydrodynamicSetup} we determine the ansatz fo the background using fluid-gravity correspondence \cite{Bhattacharyya:2008jc}. In section \ref{QfSetup} we develop the small $\Delta$ expansion and solve the fluctuation equations to first order in $\Delta$ to determine the anomalous conductivities in this order. Results of this section are applicable to a generic class of holographic theories. In section \ref{Examples} we provide three non-trivial examples and we discuss the restrictions of our results and give an outlook in section \ref{Conclusions}.

\section{Construction of the Holographic Action}\label{HolographicSetup}

To study anomaly induced transport we consider a plasma at finite temperature, electric and axial chemical potentials, and sourced by external axial and chemical magnetic fields. To include the chiral vortical effect we also need to put the theory on a slightly curved metric that we denote by $h_{\mu\nu}$. We want a plasma as close as possible to the deconfined phase of QCD. This necessitates, in the two-derivative holographic description, to include the following bulk fields corresponding to the marginal or relevant operators: the bulk metric $g_{\mu\nu}$ dual to the conserved energy-momentum tensor $T^{\mu \nu}$, a bulk gauge field $V$ dual to the vector current $J_V$, an axial gauge field $A$ the dual to the axial current $J_A$, and a bulk dilaton field $\phi$ dual to the scalar glueball operator $\text{Tr}\left(G^2 \right)$. The latter breaks the scale invariance as its VeV generates a  renormalization group invariant energy scale $\Lambda_{QCD}$. 

In addition to these fields, as explained in the Introduction, the dynamical gauge field contribution to the anomaly equation, i.e. the $a_3$ term in equation (\ref{aneq}) is realized through a bulk axion field $C_0$ that couples to the topological charge operator $\text{Tr}\left(G\wedge G \right)$ \cite{Klebanov:2002gr, Casero, Aaron, LandsteinerStuckleberg}. The boundary value of the bulk axion is proportional to the theta parameter: 
$ C_0(x,r) \to const. \times \theta + \cdots$ as $r\to \infty$. The precise form of the axion coupling\footnote{See \cite{Klebanov:2002gr, Casero, Aaron} for more general possibilities in the brane realizations.} can be inferred as follows.  Consider a space-time dependent chiral transformation $\Psi \to \exp( i \alpha_b(x) \gamma_5)\Psi$, $A_{b}\to A_{b} + d\alpha_b(x)$ in a gauge theory with massless Dirac fermions $\Psi$ coupled to an axial gauge field $A_{b}$ through a term $A_{b,\mu} J_A^\mu$. Let us, for the moment, ignore the anomalies caused by external sources, i.e. set $a_1=a_2=a_4=0$ in (\ref{aneq}). Then, the transformation of the QFT lagrangian (that originates from the fermion path integral measure, i.e. the $a_3$ term) is equivalent to a shift $\delta\theta$ in of the topological charge term $\frac{\theta N_f}{16 \pi^2} \text{Tr}\left(G_{\mu \nu} G_{\rho \sigma}   \right) \epsilon^{\mu \nu \rho \sigma}$ as we would obtain   
\begin{equation} 
{\cal L} \to  {\cal L} + \frac{1}{4} \int d^4 x \sqrt{-h}\left( \alpha_b\, \partial_\mu J_A^\mu - \left[\frac{\delta\theta N_f}{4 \pi^2}- a_3 \alpha_b\right] \text{Tr}\left(G_{\mu \nu} G_{\rho \sigma}\right) \epsilon^{\mu \nu \rho \sigma}\right)\, ,\nonumber
\end{equation}
had we also transformed the theta parameter. In the dual gravitational theory, the conservation equation follows from a bulk gauge transformation $A \to A+ d\alpha$. As the bulk gauge field and its transformation limit to the boundary values $A\to A_b$, $\alpha\to \alpha_b$ as $r\to\infty$, we then see that the correct anomaly equation (\ref{aneq}) (with $a_1=a_2=a_4=0$) would follow, if we write the bulk theory in terms of the combination $A - dC_0/Q_f$ and demand invariance under 
\begin{equation} \label{bulkinv} 
A \to A + d\alpha, \qquad C_0\to C_0 + \alpha/Q_f\, .
\end{equation}    
Here $Q_f$ is a parameter characterizing the strenght of the CP odd coupling between the axial current and the gluons, and the normalization of the bulk axion $C_0$ is such that its boundary value is equal to $\frac{Q_f N_f \theta}{4 \pi^2 a_3}$. Finally, the external anomaly terms $a_1$, $a_2$ and $a_4$ in (\ref{aneq}) is realized by including a bulk Chern-Simons term of the form $A\wedge \left[a_1 F^V\wedge F^V + a_2  F^V\wedge F^V + a_3 \text{Tr} R\wedge R\right]$ explained below.  

Having explained the necessary ingredients for a generic holographic theory for a 3+1D chiral gauge theory with anomalies we can now write down the action as
\\
\noindent
\begin{align}\label{holographicAction}
 16 \pi G_N S  &=  S_g + S_f + S_a + S_{CS} + S_{GH} + S_{ct}\, , \\
\label{glueSector}
S_{g} &= \int_{\mathcal{M}}\left[R\star 1 - \frac{1}{2} d \phi \wedge \star d \phi - V(\phi)\star 1 \right]\, ,\\
\label{flavorSector}
S_f &= - \frac{x}{2}  \int_{\mathcal{M}} \left[Z_V(\phi) F^V \wedge \star F^V + Z_A(\phi) F^A \wedge \star F^A \right]\, ,\\
\label{CPoddSector}
S_a &= -\frac{x^2 m^2}{2} \int_{\mathcal{M}}Z_0(\phi) \tilde A \wedge \star \tilde A\, , \\
\label{CSSector}
S_{CS} &=  \int_{\mathcal{M}} A \wedge \left[\kappa F^V \wedge F^V + \gamma F^A \wedge F^A + \lambda \text{Tr}\left(R \wedge R \right)\right]\, , 
\end{align}
\\
\noindent
where $16 \pi G_N=M^3_p N^2_c$ with $M_p$ the five dimension Planck scale and $S_g$, $S_f$, $S_a$, $S_{CS}$ denote the glue, flavor, axion and the Chern-Simons parts respectively\footnote{We present $S_{CS}$ in terms of $A$ instead of $\tilde A$ to make the fixing of the coefficients transparent but it has to be noted that once $S_{CT}$ is taken into account the whole action will be written in terms of $\tilde A$ \cite{LandsteinerStuckleberg}}, while $S_{GH}$ is the Gibbons-Hawking term and $S_{ct}$ denotes the counterterm action\footnote{See \cite{LandsteinerStuckleberg} for the explicit form of the counterterms.}. Labels $\{V,A\}$ stand for vector and axial bulk fields with their corresponding field strenghts $F^{V/A}$. We denote the metric of the 5D geometry by $\mathcal{G}_{M N}$ which is implicit in the action\footnote{We use uppercase latin letters for the 5D bulk, greek letters for the 4D boundary and latin indices for the 3D spatial boundary geometries.}. The axion enters the action $S_a$ in the gauge invariant combination
\begin{equation}\label{Atilde}
\tilde A \equiv A - \frac{dC_0}{Q_f}\equiv A - d \mathfrak{a}\, .
\end{equation} 
Thus $S_a$ provides both a kinetic term for the axion and a St\"uckelberg mass term for the axial gauge field. 

We allow for arbitrary potentials $V(\phi)$, $Z_V(\phi)$ and $Z_A(\phi)$ for the dilaton $\phi$ and its coupling to the vector and axial gauge fields. The potentials are normalized such that
\begin{equation}
\lim_{r \rightarrow \infty} Z_A(\phi) = \lim_{r \rightarrow \infty} Z_V(\phi) = \lim_{r \rightarrow \infty} Z_0(\phi) = 1,
\end{equation}
where $r$ denotes the holographic radial coordinate and the AdS like boundary is located at $r \rightarrow \infty$. 

The coefficient $x=\frac{N_f}{N_c}$ is the Veneziano parameter defined in (\ref{Veneziano}). The scaling with $x$ of each term in the action can be deduced from the original string action as discussed in \cite{thetaKiritsis}. The coupling parameter $Q_f$ is related to the Veneziano parameter via 
\begin{equation}
\label{Qf}
Q_f = mx\, ,
\end{equation}
where $m$ is a constant with mass dimension that in principle can be derived from the original string theory model. In this work we consider $Q_f$ as a tunable parameter. We develop a series expansion in $Q_f$, hence assume $Q_f$ small.  In field theory this corresponds to weak CP odd coupling of the gluons to the fermion. It is important to note that small $Q_f$ does not necessarily imply small $x$ and we consider $x$ to be a free parameter.

Finally, in order to fix the Chern-Simons coefficients $\kappa$, $\gamma$ and $\lambda$ in (\ref{CSSector}) we consider the variation of the axial gauge field $\delta A = d\alpha$ under which
\begin{equation}\label{variationBulk}
\delta S= \frac{1}{16 \pi G} \int \alpha \left[\kappa F^V \wedge F^V + \gamma F^B \wedge F^B + \lambda \text{Tr}\left(R \wedge R \right) \right].
\end{equation}
Comparing \eqref{variationBulk} with \eqref{aneq}  we find
\begin{equation}\label{CScoefficients}
\kappa = - 16 \pi G_N a_1, \quad \quad \gamma = - 16 \pi G_N a_2, \quad \quad \lambda= - 16 \pi G_N a_4.
\end{equation}

The equations of motion obtained from the variation of the action \eqref{holographicAction} read as follows. For the dilaton and the axion fields we obtain
\begin{equation}\label{dilatonEquation}
d\star d \phi = \partial_\phi V \star 1 + x \frac{\partial_\phi Z_V}{2} F^V \wedge \star F^V + x\frac{\partial_\phi Z_A}{2} F^A \wedge \star F^A + Q^2_f\frac{\partial_\phi Z_0}{2} \tilde A \wedge \star \tilde A, 
\end{equation}
\begin{equation}\label{axionEquation}
d\left(Z_0 \star \tilde A \right) = 0,
\end{equation}
For the gauge fields we have, 
\begin{equation}\label{vectorEquation}
d \left(x Z_V \star F^V - 2 \kappa A \wedge F^V \right) = 0,
\end{equation}

\begin{equation}\label{axialEquation}
d\left[ x Z_A \star F^A - \kappa V \wedge F^V - 3 \gamma A \wedge F^A  - \lambda \text{Tr}\left(\omega \wedge d \omega + \frac{2}{3} \omega \wedge \omega \wedge \omega \right) \right] = - Q^2_f Z_0 \star \tilde A\, .
\end{equation}
Finally the Einstein's equations are 
\begin{equation}\label{Einstein}
\begin{split}
R_{M N}= \frac{1}{2} \partial_M \phi \partial_N \phi + \frac{Q^2_f  Z_0}{2} \tilde A_M \tilde A_N 
+ \frac{V}{3} \mathcal{G}_{MN}\\  + \frac{x Z_V}{2} \left(F^V_{MP} F^{V,P}_{\hphantom{V,P}N}- \frac{1}{6} \mathcal{G}_{MN} F^V_{PS} F^{V,PS} \right) \\
+\frac{ x Z_A}{2} \left(F^A_{MP} F^{A,P}_{\hphantom{A,P}N}- \frac{1}{6} \mathcal{G}_{MN} F^A_{PS} F^{A,PS} \right) \\ + \frac{\lambda}{2} \left[\nabla_L \left(\Sigma^L_{\hphantom{L}(NM)} - \frac{1}{3}\mathcal{G}_{MN}  \Sigma^{L\hphantom{M}M}_{\hphantom{L}M} \right) \right]\, ,
\end{split}
\end{equation}
where $\omega$ is the spin connetion and $\Sigma^L_{\hphantom{L}MN}$ is defined as 
\begin{equation}
\Sigma^L_{\hphantom{L}MN} = - \mathcal{G}_{M P_1} \epsilon^{P_1 P_2 P_3 P_4 P_5} F^A_{P_2 P_3} R_{P_4 P_5 N}^{\hphantom{P_4 P_5 N} L}\, .
\end{equation}
Here $\epsilon^{MNPQR}$ denotes the 5D Levi-Civita tensor. 

As explained in the Introduction our purpose is to calculate the one point function of the vector current and read off the anomalous conductivities. The holographic prescription for this one point function is 
\\
\begin{equation}\label{OnePoint}
\langle J^\nu_V \rangle  = \frac{1}{16 \pi G_N} \lim_{ r \rightarrow \infty} \left[- x \sqrt{-G} Z_V  F^{V,r\nu} + 2 \kappa \tilde \epsilon^{\nu \mu \rho \sigma} A_\mu F^V_{\rho \sigma} \right],
\end{equation}
\\
\noindent
where $\tilde \epsilon^{\mu \nu \rho \sigma}$ is the Levi-Civita symbol. This form of the one point function includes holographic renormalization \cite{LandsteinerStuckleberg}. 

\section{Background ansatz}\label{HydrodynamicSetup}

\subsection{Background at equilibrium} 

We consider a general ansatz for the background obtained by the hydrodynamic setting we want to describe through the fluid-gravity correspondence \cite{Bhattacharyya:2008jc}.  First consider an equilibrium configuration characterized by a (boundary) background metric $h_{\mu \nu}$, a four velocity $u^\mu$ normalized as\footnote{Boundary greek indices $\{\mu,\nu, ... \}$ are raised and lowered by the metric $h_{\mu \nu}$.} $u_\nu u_\mu h^{\mu \nu} = -1$, a chemical potential $\mu$, an axial chemical potential\footnote{As the axial current is non-conserved $\mu_5$ should be thought of as a coupling in the Hamiltonian rather than a true chemical potential\cite{LandsteinerStuckleberg}.} $\mu_5$, an equilibrium temperature $T$, an external vector and axial sources $\tilde v$ and $\tilde a$. At this point all fields are taken to be constant.

To represent magnetic interactions the vector and axial sources are taken transverse to the direction of propagation,  i.e. $u^\mu \tilde v_\mu=0$ and $u^\mu \tilde a_\mu=0$. It is possible to use the four velocity to decompose any tensor structure into a projection along the propagation and transverse to it. The transverse projector is
\\
\begin{equation}
\Delta_{\mu \nu}= h_{\mu \nu} + u_\mu u_\nu,
\end{equation}
\\
\noindent
which satisfies $\Delta_{\mu \nu} u^\mu=0$ and $\Delta^{\mu \rho} \Delta_{\rho \nu}=\Delta^\mu_\nu$. We can then write down the following ansatz for the metric, the gauge fields and the scalars\footnote{We do not use Eddington-Finkelstein coordinates unlike what is usually done in the fluid gravity correspondence. Our calculations will be at exactly zero frequency and regularity at the horizon is enough to determine the boundary conditions.} 
\\
\begin{equation}\label{ansatz0Metric}
ds^2= \frac{dr^2}{g(r) r^2} + r^2 \left[- f(r)u_\mu u_\nu +  \Delta_{\mu \nu}  + \mathfrak{A}(r) u_\mu \tilde a_\nu + \mathscr{A}(r)\tilde a_\mu \tilde a_\nu   \right] dx^\mu dx^\nu, 
\end{equation}
\begin{equation}\label{ansatz0Vector}
V= -V_t(r) u + \mathcal{V}(r) \tilde a +  \tilde v,
\end{equation}
\begin{equation}\label{ansatz0Axial}
A= -A_t(r) u + \mathcal{A}(r) \tilde a,
\end{equation}
\begin{equation}
\phi=\phi(r), \quad \quad \mathfrak{a}=\mathfrak{a}(r),
\end{equation}
\\
\noindent
where $u=u_\mu dx^\mu$, $\tilde v=v_\mu dx^\mu$ and $\tilde a=a_\mu dx^\mu$. The functions $f(r)$ and $g(r)$ are the blackening factors, the functions $A_t(r)$ and $V_t(r)$ determine the chemical potentials on the boundary theory as explained below and $\mathcal{V}(r),\mathcal{A}(r),\mathscr{A}(r)$ and $\mathfrak{A}(r)$ characterize back reaction of the axial source to the vector and axial bulk gauge fields and the metric. This can be contrasted with the massless case where a constant gauge source does back react onto the rest of the background. 
Although we will only be interested in the linear response to the sources we will keep the non-linear terms---that naturally appear in the Ansatz---for consistency and generality, until the end of our calculations. 

\vspace{2mm}
For  \eqref{ansatz0Metric} to be asymptotically AdS with the boundary metric $h_{\mu \nu}$ we require
\\
\begin{equation}\label{boundaryAdS}
\begin{split}
\lim_{r \rightarrow \infty} f(r) =1, \quad \quad \lim_{r \rightarrow \infty} g(r)=1, \quad \quad \lim_{r \rightarrow \infty} \mathfrak{A}(r) =0, \quad \quad \lim_{r \rightarrow \infty} \mathscr{A}(r)=0.
\end{split}
\end{equation}
\\
\noindent
We also require a non-extremal horizon at $r_h$ 
\\
\begin{equation}
\begin{split}
f(r) \sim f_1 (r-r_h) + f_2 (r-r_h)^2 + ... \\ g(r) \sim g_1 (r-r_h) + g_2 (r-r_h)^2 +...
\end{split}
\end{equation}
\\
\noindent 
and regularity at $r_h$ for all the other background functions.
We read off the temperature from the  horizon data as 
\\
\begin{equation}
T= \frac{r^2_h}{4 \pi} \sqrt{f_1 g_1}.
\end{equation}
\\
\noindent
Boundary asymptotics of the gauge fields are (see e.g. \cite{LandsteinerStuckleberg})
\\
\begin{equation}
\label{gfasymp}
\begin{split}
\lim_{ r \rightarrow \infty}V(r) &\sim c_1 + \frac{c_2}{r^2},  \\
\lim_{r \rightarrow \infty} A(r) &\sim r^\Delta c_3+ \frac{c_4}{r^{2-\Delta}},
\end{split}
\end{equation}
\\
\noindent
where $c_i$ are constant one-forms. The power $\Delta$ corresponds  to the anomalous dimension of the axial current on the boundary, given in terms of the parameters in the action as  
\\
\begin{equation}
\label{Delta}
\Delta= \sqrt{1 + \frac{Q^2_f}{x}} - 1 = \sqrt{1 + m^2\, x} - 1 \, .
\end{equation}
\\
\noindent
From the powers in the normalizable modes in (\ref{gfasymp}) one reada the scaling dimension of the dual vector and axial currents as dim[$J_V$]=3 and dim[$J_A$]=$3+\Delta$. To avoid axial current becomes irrelevant in the IR we need to require $\Delta<1$. 

\vspace{2mm}
The chemical potentials of the boundary field are given in terms of the gauge invariant expressions below
\\
\begin{equation}\label{boundaryBackground}
\int^\infty_{r_h} dr  V'_t(r) = \mu, \quad \quad  \lim_{R \rightarrow \infty} \left(\frac{R}{r_h}\right)^{-\Delta} \int^R_{r_h}dr A'_t (r) = \mu_5\, ,
\end{equation}
\\
\noindent
where prime denotes a radial derivative. Using the regularity of the gauge fields at the horizon\footnote{See \cite{LandsteinerReview,LandsteinerStuckleberg} for a careful discussion on the regularity of the gauge fields at the horizon and different ways to introduce the chemical potentials in the bulk dual.} 
\begin{equation}
V_t(r_h)= A_t(r_h)=0\, ,
\end{equation}
equations (\ref{boundaryBackground}) imply 
\begin{equation}
\lim_{r \rightarrow \infty} V_t(r) = \mu\, , \qquad 
\lim_{r \rightarrow \infty} \left[\frac{r}{r_h}\right]^{-\Delta} A_t(r) = \mu_5\, .
\end{equation}

\subsection{Fluctuations}
\label{flucs}

To study fluctuations around the equilibrium configuration \eqref{ansatz0Metric}-\eqref{ansatz0Axial} we promote the background fields to slowly variating functions of the coordinates $\{x^\mu\}$ that remain static with respect to a timelike Killing vector $\xi^\mu$, namely $\mathcal{L}_\xi \Phi=0$ for any field $\Phi$. Under these conditions the Ansatz \eqref{ansatz0Metric}-\eqref{ansatz0Axial} will no longer be a solution to the equations of motion but it can be corrected order by order in a derivative expansion\cite{Bhattacharyya:2008jc}, 
\\
\begin{equation}
\Phi(x)=\sum \epsilon^n \Phi^{(n)}(x), \nonumber
\end{equation}
\\
\noindent
where $n$ denotes the number of derivatives and $\epsilon$ is a book keeping parameter to track the order in the derivative expansion. The ansatz \eqref{ansatz0Metric}-\eqref{ansatz0Axial}, with fields promoted to functions of the boundary coordinates $x^\mu$ correspond to the zeroth order solution $\Phi^{(0)}$. At this order we require  the four velocity $u^\mu$ be proportional to the constant Killing vector $\xi^\mu$. Only the corrections up to first order in $\epsilon$ will be relevant in our calculations below, and $\mathcal{O}\left( \partial^2\right)$  contributions will be disregarded. 

\vspace{4mm}
The full hydrodynamic Ansatz is then given by\footnote{A more general Ansatz reads  \cite{Erdmenger,Minwalla}
\begin{eqnarray}
ds^2 &=&\frac{dr^2}{gr^2} + r^2\left[-f u_\mu u_\nu + \Delta_{\mu \nu} + \mathfrak{A}u_\mu \tilde a_\nu   + \mathscr{A}  a_\mu \tilde a_\nu \right] + \Pi_{\mu \nu} dx^\mu dx^\nu + L_\mu u_\nu dx^\mu dx^\nu, \nonumber \\
A&=&-A_t u_\mu dx^\mu + A^{\bot}_\mu dx^\mu,\quad \quad 
V=-V_t u_\mu dx^\mu + V^{\bot}_\mu dx^\mu, \nonumber
\end{eqnarray}
where $\Pi_{\mu},L_\mu,V^{\bot}_\mu$ and $A^{\bot}_\mu$ are all transverse to the fluid velocity. When these terms are evaluated on an equilibrium configuration only the terms shown in \eqref{metricAnsatz} remain. } 
\\
\begin{eqnarray}\label{metricAnsatz}
ds^2 &=&\frac{dr^2}{g r^2} +  r^2 \left[-f u_\mu u_\nu + \Delta_{\mu \nu} + \mathfrak{A} u_\mu \tilde a_\nu  + \mathscr{A} \tilde a_\mu \tilde a_\nu\right] dx^\mu dx^\nu \nonumber\\ 
{}&&+ r^2 \left[ \gamma_I u_\mu  B^I_\nu + \kappa_I   \tilde a_\mu B^I_\nu \right] dx^\mu dx^\nu  + \mathcal{O}\left(\partial^2 \right), \\
\label{VAnsatz}
V &=&-V_t u + \mathcal{V} \tilde a + \tilde v + \beta_I B^{I} + \mathcal{O}\left(\partial^2 \right)\, ,\\
\label{AAnsatz}
A &=& -\tilde A_t u + \mathcal{A} \tilde a + \alpha_I B^I + \mathcal{O}\left(\partial^2 \right)\, , \\
\label{dilatonAnsatz}
\phi &=&\phi(r,x), \quad \quad \mathfrak{a}=\mathfrak{a}(x,r)\, ,
\end{eqnarray}
\\
\noindent
where the index $I$ runs over all the magnetic sources and we fix the thermodynamic sources to constant values. The metric and gauge field solutions are ordered such that the boundary coordinates only appear through the sources while all the dependence on the radial coordinate\footnote{Due to the mass term the blackening functions are corrected by $\tilde a^2$. This plays no role in the linear response regime considered in this paper but might become relevant in another holographic context.} is in the functions $f,g, \tilde A_t, \mu_t, V_t$ ,$\mathcal{V}, \mathcal{A},\mathfrak{A}, \mathscr{A},\alpha, \beta$, $\kappa$ and $\gamma$. To ensure regularity of the Ricci scalar at the horizon the functions $\gamma_I$ and $\mathfrak{A}$ should satisfy \cite{GursoyTarrio} 
\\
\begin{equation}
\gamma_I(r_h)=\mathfrak{A}(r_h)=0\, .
\end{equation}
\\
\noindent
Finally, the magnetic field forms $B^{V/A}=B^{V/A}_\mu dx^\mu$ and the vorticity form $B^\omega=\omega=\omega_\mu dx^\mu$ are defined by
\\
\begin{equation}
\begin{split}
B^{V,\mu}=\epsilon^{\mu \nu \rho \sigma} u_\nu \partial_\rho \tilde v_\sigma  \quad \quad B^{A,\mu}=\epsilon^{\mu \nu \rho \sigma} u_\nu \partial_\rho \tilde a_\sigma  \quad \quad \omega^\mu= \epsilon^{\mu \nu \rho \sigma} u_\nu \partial_\rho u_\sigma \, .
\end{split}
\end{equation}
\\
\noindent
In above we set the axial gauge $V_r=0$ and $A_r=0$\footnote{We also moved the functional dependence of $A_r$ is to $\mathfrak{a}$.}.

\section{Solution to fluctuations at small $\Delta$} 
\label{QfSetup}

In this section we solve the background equations of motion that we derived in section \ref{HolographicSetup} on the Ansatz of section \ref{flucs} perturbatively in the parameter $Q_f$. In particular we will be interested in the solution up to ${\cal O}(Q_f^2)$. Assumption of small $Q_f$ corresponds to small anomalous dimension $\Delta$, c.f. equation (\ref{Delta}), hence, a weak contribution of the mixed gauge-global axial anomaly in internal Feynman diagrams. Recalling the derivative expansion in the hydrodynamic picture, that we denoted by $\epsilon$ in section \ref{flucs}, any field $\Phi$ formally admits a double expansion of the form
\\
\begin{equation}
\Phi(x,r) = \sum_m \sum_n Q^{(m)}_f \Phi^{(n,m)} \epsilon^n.
\end{equation}
\\
\noindent
We only consider contributions up to $n=1$ in the hydrodynamic expansion, as this is sufficient for our purpose to compute the conductivities. Similarly, we keep only terms up to $m=2$ in the $Q_f$ expansion. We note that, thi series actually start at $m=2$ as the mass term in (\ref{CPoddSector}) first appears at this order. Furthermore, the two expansions are of different nature therefore they do not mix. 

The external hydrodynamic and thermodynamic sources $\{u,\tilde a, \tilde v, \mu, \mu_5,T \}$ are taken $\mathcal{O}(1)$. The axion $\mathfrak{a}$ is assumed to be at least of $\mathcal{O}\left(1\right)$ while the  functions $\{\mathcal{V},\mathcal{A}, \mathfrak{A}, \mathscr{A}, \kappa \}$ are  
\\
\begin{equation}
\begin{split}
\mathcal{V}  &\sim  \mathfrak{A} \sim \mathscr{A} \sim  \kappa \sim \mathcal{O}\left(Q^2_f \right), \\
\mathcal{A} &= 1+ \mathcal{O}\left(Q^2_f \right).
\end{split}
\end{equation}
\\
\noindent
Scaling properties of these functions then allow the $\mathcal{O}\left(\partial^{0} \right)$ metric components $\mathcal{G}_{\mu \nu}$ to split into a part $g_{\mu \nu}$ with the lowest term $\mathcal{O}(Q_f^0)$ and a part $\mathfrak{G}$ with the lowest term $\mathcal{O}\left(Q^2_f\right)$
\\
\begin{equation}
\begin{split}
g_{\mu \nu}= r^2 \left[-f u_\mu u_\nu + \Delta_{\mu \nu} \right], \quad \quad \mathfrak{G}_{\mu \nu}=r^2 \left[\mathfrak{A} u_{(\mu} \tilde a_{\nu)} + \mathscr{A} \tilde a_\mu \tilde a_\nu \right].
\end{split}
\end{equation}
\\
\noindent
As we keep terms only up to $Q^2_f$, we can write the bulk metric and its inverse as 
\\
\begin{equation}
\mathcal{G}_{MN}= \begin{bmatrix}
  \frac{1}{r^2 g} & 0 \\
  0 & g_{\mu \nu} + \mathfrak{G}_{\mu \nu} + \mathcal{O}\left(\partial^2 \right)
 \end{bmatrix}, \quad 
 \mathcal{G}^{MN} = \begin{bmatrix}
 r^2 g & 0 \\
 0 & g^{\mu \nu} + \mathfrak{G}^{\mu \nu} + \mathcal{O}\left(\partial^2,Q^4_f \right)
 \end{bmatrix}\, ,
\end{equation}
\\
\noindent
with $g^{\mu \nu}$ and $\mathfrak{G}^{\mu \nu}$ given by 
\\
\begin{equation}
\begin{split}
g^{\mu \nu}= \frac{1}{r^2}\left[- \frac{u^\mu u^\nu}{f} + \Delta^{\mu \nu} \right], \quad \quad 
\mathfrak{G}^{\mu \nu} = -\frac{1}{r^2}\left[ \mathscr{A} \tilde a^\mu \tilde a^\nu +\frac{\mathfrak{A}}{f} u^{(\mu} \tilde a^{\nu)} \right]\, .
\end{split}
\end{equation}
\\
\noindent
Two comments are in order. The form of the inverse metric is exact at all orders in $Q_f$ in the linear response regime. Moreover, the $Q_f$ expansion of the anomalous dimension $\Delta$  is given by $\Delta = \frac{Q^2_f}{2x} + \mathcal{O}\left(Q^4_f\right)$.

Below we discuss the equations of motion and their solutions at $\mathcal{O}\left(\partial^0\right)$  and  $\mathcal{O}\left(\partial\right)$, one by one.

\subsection{The dilaton} 

\noindent The schematic equation of motion of the dilaton reads

\begin{equation}\label{axionOrder0}
\begin{split}
\frac{1}{\sqrt{-G}} \partial_r \left[ \sqrt{-G} r^2 g \phi' \right] = \partial_\phi V + \Phi\left( Q^2_f \tilde a^2, \mathcal{A}'\partial, \mathcal{V}'\partial  \right) + \mathcal{O}\left(\partial^2 \right) ,
\end{split}
\end{equation}
\\
\noindent
where $\Phi$ is some function of the indicated variables. We note that, in the linear response regime, the dilaton only couples at $\mathcal{O}\left(\partial^2 \right)$ hence its hydrodynamic corrections will play no role in what follows.

\subsection{The axion} 

\noindent The equation of motion \eqref{axionEquation} for the axion can be expanded as
\\
\begin{equation}\label{axionOrder0}
\begin{split}
\partial_r  \left(\sqrt{-G} Z_0 r^2 g \mathfrak{a}'  \right)=\frac{Z_0 \sqrt{-G}}{r^2} \left[- \left( \frac{\mathfrak{A} \mathcal{A}}{2f} \right) u^\mu \partial_\mu \tilde a^2  -\left( \frac{2A_t+ \mathfrak{A} \mathcal{A} \tilde a^2}{2f} \right) \partial_\mu u^\mu \right. \\  \left. + \left( \mathcal{A} - \frac{A_t \mathfrak{A}}{2f} - \mathcal{A} \mathscr{A} \tilde a^2\right)  \partial_\mu \tilde a^\mu  - \mathcal{A}\mathscr{A} \tilde a^\mu \partial_\mu \tilde a^2 \right] + \mathcal{O}\left( \partial^2 \right)\, ,
\end{split}
\end{equation}
\\
\noindent
which, in turn can easily be integrated as,
\\
\begin{equation}\label{axionSolution}
\begin{split}
\mathfrak{a} = \mathfrak{a}_0 + \int_{r_h}^{r}dr' \frac{\left( \partial_\mu \tilde a^\mu \right) }{\sqrt{-G}Z_0 r^2 g}\left[  \int_{r_h}^{r'} dy \frac{Z_0 \sqrt{-G}}{r^2} \left(  \mathcal{A} - \frac{A_t \mathfrak{A}}{2f} - \mathcal{A}\mathscr{A}\right) \right] \\ -  \int_{r_h}^{r}dr' \frac{\left(\tilde a^\mu \partial_\mu \tilde a^2 \right) }{\sqrt{-G}Z_0 r^2 g}\left[  \int_{r_h}^{r'} dy \frac{Z_0 \sqrt{-G}}{r^2} \mathcal{A} \mathscr{A} \right] + \mathcal{O}\left(\partial^2 \right).
\end{split}
\end{equation}
\\
\noindent
Here the non-normalizable mode $\mathfrak{a}_0(x)$ is independent of the radial coordinate $r$, and it corresponds to a space-time dependent $\theta$-term in QCD.  More precisely, from equation (\ref{bulkinv}), (\ref{Atilde}) and the discussion in between, we have the constant (x-independent) part of $\mathfrak{a}_0$ equals $N_f \theta/4\pi^2a_3$.  
To derive (\ref{axionSolution}) from  (\ref{axionOrder0}) we used $g(r_h)=0$, and 
\\
\begin{equation}
\label{conds1}
\partial_\mu u^\mu=\mathcal{O}\left(\partial^2\right)\, , \qquad u^\mu \partial_\mu \tilde a = 0 \, .
\end{equation}
\\
\noindent
The first one is just incompressibility of the fluid at linear order and the second one is best understood in the rest frame where it means that there are no electric components in the external gauge field $\tilde a$. 

The normalizable mode given by the integral expression in \eqref{axionSolution} is identified with the expectation value $\langle  \text{Tr}\left(G\wedge G \right) \rangle$. This implies that $\langle  \text{Tr}\left(G\wedge G \right) \rangle \sim \mathcal{O}\left(\partial \right)$. However, as discussed in \cite{generalCurrent}, validity of the hydrodynamic expansion requires that the ${\cal O}(\partial)$ contribution to  $\langle  \text{Tr}\left(G\wedge G \right) \rangle$ vanish, to prevent any contribution of gluons to entropy production. This means that we should require 
\\
\begin{equation}
\label{conds2}
\partial_\mu \tilde a^\mu=0\, , \qquad \tilde a^\mu \partial_\mu \tilde a^2 = 0 \, .
\end{equation}
\\
\noindent
While the first condition is again natural since we require the source $\tilde a$ be magnetic-like hence divergenceless, the second one imposes a non-trivial restriction on the external sources we consider in this paper\footnote{Note that a canonical example of a constant (axial) magnetic field e.g. $\tilde a^\mu = (0, -yB^5/2, x B^5/2,0)$ satisfies it.}. This means, 
\\
\begin{equation}
\label{conds3}
\mathfrak{a} = \mathfrak{a}_0 + \mathcal{O}\left(\partial^2\right) \, .
\end{equation}
\\
\noindent
That is, consistency with positive local entropy production requires axion to be constant up to second order in derivatives. 

\subsection{The vector field}

\noindent The vector field equation of motion \eqref{vectorEquation} at the zeroth order in the derivative expansion reads
\vspace{2mm}
\begin{equation}
\partial_r \left[ \frac{\sqrt{-G} Z_V g V'_t}{f} \left(u^\nu + \left(\frac{\mathfrak{A}}{2} - \frac{\mathcal{V}' f}{V'_t} \right) a^\nu  \right) \right] = \mathcal{O}\left(Q^4_f \right),
\end{equation}
\\
\noindent
which results in translating in an expression for the conserved charge $Q$, and a relationship between the background functions $\mathfrak{A}$ and $\mathcal{V}$
\vspace{2mm}
\begin{equation}\label{vectorConditions}
\begin{split}
Q= \frac{\sqrt{-G} Z_V g V'_t}{f},   \quad \quad  Q'= 0, \quad \quad 
\mathcal{V}'= \frac{V'_t \mathfrak{A}}{2f}.
\end{split}
\vspace{2mm}
\end{equation}
\noindent
To study the vector equation  at $\mathcal{O}\left(\partial\right)$ we need the CS form $F^V \wedge F^A$ and the membrane current $\bar J^\nu=\sqrt{-\mathcal{G}}x Z_VF^{V,r\nu}$ up to this order in the derivative expansion. We find 
\vspace{2mm}
\begin{eqnarray}\label{CSV}
F^V &=& - V_t du + \mathcal{V} d \tilde a + d\tilde v + \left(-V'_t u + \mathcal{V}' a \right) \wedge dr + \mathcal{O}\left( \partial^2 \right), \\
\label{CSA}
F^A &=& - A_t du + \mathcal{A} d \tilde a + \left(-A'_t u + \mathcal{A}' \tilde a \right) \wedge dr + \mathcal{O}\left( \partial^2 \right). 
\vspace{2mm}
\end{eqnarray}
\noindent
The membrane current can schematically be expressed as

\vspace{2mm}
\begin{equation}\label{membraneCurrent1}
\bar J^\nu = x Qu^\nu + \sqrt{-\mathcal{G}} x Z_V g \left[ \bar  J_{1I}(r) \left(B^I \cdot \tilde a \right) u^\nu + \bar J_{2I}(r)  \left(B_I \cdot \tilde a \right) \tilde a^\nu + \bar J_{3I}(r) B^{I\nu}   \right] + \mathcal{O}\left( \partial^2 \right)\, , 
\end{equation}

\vspace{2mm}
\noindent
where we defined the functions $\bar J_{1I}$ , $\bar J_{2I}$ and $\bar J_{3I}$ 

\vspace{2mm}
\begin{eqnarray}
\bar J_{1I} (r) &=& \frac{V'_t \gamma_I \mathfrak{A}}{2 f^2} - \frac{\mathcal{V}' \gamma_I}{2f} - \frac{\beta'_I \mathfrak{A}}{f} + \mathcal{O}\left(Q^4_f \right), \\
\bar J_{2I}(r) &=& \frac{V'_t \gamma_I \mathscr{A}}{2f}  - \beta'_I \mathscr{A} + \mathcal{O}\left(Q^4_f \right), \\
\bar J_{3I}(r) &=& - \frac{V'_t \gamma_I}{2f} + \beta'_I + \mathcal{O}\left(Q^4_f \right).
\end{eqnarray}

\vspace{2mm}\noindent
What is relevant for our calculation below is that $\bar J^\nu(r_h)= x Q u^\nu$ which follows from the boundary conditions $\gamma_I(r_h)=\mathfrak{A}(r_h)=g(r_h)=f(r_h)=0$. Using \eqref{CSV}, \eqref{CSA} and \eqref{membraneCurrent1} the vector equation up to $\mathcal{O}\left(\partial \right)$ can be integrated out into

\vspace{2mm}
\begin{equation}\label{membraneCurrent}
\begin{split}
\bar J^\nu &=  2\kappa \tilde \epsilon^{\nu \mu \rho \sigma} \left[  A_t V_t u_\mu \partial_\rho u_\sigma - A_t u_\mu \partial_\rho  \tilde v_\sigma + \mathcal{A} \tilde a_\mu \partial_\rho \tilde v_\sigma  + \mathcal{A} \mathcal{V} \tilde a_\mu \partial_\rho \tilde a_\sigma\right]  \\ 
{}&- 2\kappa \tilde \epsilon^{\nu \mu \rho \sigma} \int^r_{r_h} dr \left[ \left(A_t \mathcal{V}' + \mathcal{A}'V_t \right) \tilde a_\mu \partial_\rho u_\sigma + \left(A'_t \mathcal{V} + \mathcal{A} V'_t \right) u_\mu \partial_\rho \tilde a_\sigma  \right]  \\ {}&-2 \kappa \mathcal{A}(r_h)  \tilde \epsilon^{\nu \mu \rho \sigma} \tilde a_\mu \partial_\rho \tilde v_\sigma  + x Q u^\nu + \mathcal{O}\left(\partial^2 \right)\, .
\end{split}
\end{equation}

\vspace{2mm}
\noindent where $\tilde \epsilon^{\mu \nu \alpha \beta}$ denotes the Levi-Civita symbol. The bulk version of the consistent vector current \eqref{OnePoint} is obtained by shifting $\bar J$ with the Chern-Simons current as\footnote{$\tilde J^\nu$ itself gives the covariant current.} 

\vspace{2mm}
\begin{equation}
\tilde J^\nu _V \equiv  - \tilde J^\nu + 2 \kappa \epsilon^{\nu \mu \rho \sigma} A_\mu F^V_{\rho \sigma},
\end{equation}

\vspace{2mm}
\noindent
and extracting its boundary value 

\vspace{2mm}
\begin{equation}\label{16piG}
\lim_{r \rightarrow \infty} \tilde J^\nu_V =16 \pi G_N \langle J^\nu_V \rangle.
\end{equation}

\vspace{2mm}\noindent
From \eqref{membraneCurrent} the holographic vector current is found to be given by

\vspace{2mm}
\begin{equation}\label{vectorCurrentOne}
\begin{split}
\tilde J^\nu_V &= -Q x u^\nu + 2 \kappa \mathcal{A}(r_h) \tilde \epsilon^{\nu \mu \rho \sigma} \tilde a_\mu \partial_\rho \tilde v_\sigma 
\\ {}&- 2 \kappa \tilde \epsilon^{\nu \mu \rho \sigma} \int^r_{r_h} dr \left[ A_t \mathcal{V}' - \mathcal{A} V'_t \right]\left[u_\mu \partial_\rho \tilde a_\sigma - \tilde a_\mu \partial_\rho u_\sigma  \right] + \mathcal{O}\left(\partial^2 \right).
\end{split}
\end{equation}

\vspace{2mm}\noindent
The holographic current \eqref{vectorCurrentOne} contrary to the membrane current in \eqref{membraneCurrent} has a well defined $r \rightarrow \infty$ limit. Equation \eqref{vectorCurrentOne} together with \eqref{vectorConditions} constitute the main results of this section.

\subsection{The axial field} 

\noindent
The Maxwell equation for the axial gauge field at $\mathcal{O}\left(\partial^0\right)$ reads 

\vspace{2mm}
\begin{equation}
\partial_r\left[\frac{\sqrt{-\mathcal{G}}Z_A  g A'_t}{f} \left[ u^\nu + \left(\frac{\mathfrak{A}}{2} - \frac{\mathcal{A}' f}{A'_t} \right) a^\nu  \right] \right]= \frac{Q^2_f}{x}\frac{\sqrt{-\mathcal{G}}Z_0}{r^2}\left[\frac{A_t u^\nu}{f} - \mathcal{A} \tilde a^\nu  \right] + \mathcal{O}\left(Q^4_f \right),
\end{equation}

\vspace{2mm}\noindent
This equation decouples into the equations for the background

\vspace{2mm}
\begin{equation}\label{axialCharges}
 Q_5 = \frac{\sqrt{-\mathcal{G}} Z_A g A'_t}{f}, \quad \quad Q'_5 =  \frac{Q^2_f}{x} \frac{\sqrt{-\mathcal{G}} Z_0 A_t}{r^2 f},  
\end{equation}

\vspace{2mm}\noindent
and a coupled equation for $\mathcal{A}$ and $\mathfrak{A}$ that can, alternatively, be written in terms of $\mathcal{V}$ as 

\vspace{2mm}
\begin{equation}\label{Acal}
\partial_r \left[ Q_5 f \left(\frac{\mathcal{A}'}{A'_t} - \frac{\mathcal{V}'}{V'_t} \right) \right]= \frac{Q'_5 f \mathcal{A}} {A_t} + \mathcal{O}\left(Q^4_f \right).
\end{equation}

\vspace{2mm}
\noindent Equations \eqref{axialCharges} and \eqref{Acal} are the main results that is used below.

\subsection{Einstein equations}

At zeroth order in the derivative expansion Einstein equations can be projected into the radial direction, in a direction along the velocity, and a direction orthogonal to the velocity. By combining three of the four independent equations, as shown in appendix \ref{appendixB}, the following relation is found

\begin{equation}\label{gravitationalConstant}
\left[ r^5 f' \sqrt{\frac{g}{f}} - x \left(A_t Q_5 + Q V_t \right) \right]' = \hat \rho \left( \tilde a^2 Q^2_f, \partial \right),
\end{equation}

\noindent where $\hat \rho$ is some function. Equation \eqref{gravitationalConstant} implies that up to $\mathcal{O}\left(Q_f \right)$, or alternatively in the linear response regime, there exists a conserved gravitational charge. We note that for the set of equations \eqref{tangentPro}-\eqref{velocityPro} to be consistent, both $f$ and $g$ should receive an $\mathcal{O}\left(\tilde a^2 Q^2_f \right)$ correction. This will have no consequence in the analysis of this paper as we are only interested in linear response.
\vspace{2mm}
A similar analysis for the projection along $u^\mu \tilde a^\nu$ also yields a total derivative   

\vspace{2mm}
\begin{equation}\label{gravitationalVariationConstant}
\begin{split}
 \left[ r^{5}\sqrt{\frac{g}{f}} \frac{\mathfrak{A'}}{2} -x \left( Q_5 \mathcal{A} + Q \mathcal{V} \right) \right]' = \mathcal{O}\left(Q^4_f, \partial \right).
\end{split}
\end{equation}

\vspace{2mm}\noindent
It is possible to integrate equation \eqref{gravitationalVariationConstant} twice by making use of equations \eqref{gravitationalConstant} and \eqref{Acal} together with their universal near horizon behavior, all in all, resulting  in the following simpler formula

\vspace{2mm}
\begin{equation}\label{Vcal}
\partial_r \left( \frac{\mathcal{V}'}{V'_t}\right) = \frac{x Q_5 A_t}{r^5 \sqrt{fg}} \left( \frac{\mathcal{A}'}{A'_t} - \frac{\mathcal{V}'}{V'_t} \right) + \mathcal{O}\left(Q^4_f\right)\, .
\end{equation}

\vspace{2mm}\noindent 
which is the main result of this analysis. This equation, together with \eqref{Acal} yields a solution to $\mathcal{A}$ and $\mathcal{V}$ up to $\mathcal{O}\left(Q^2_f\right)$. We emphasize that both \eqref{Vcal} and \eqref{Acal} become exact in the linear response regime.

\subsection{The $\mathcal{A}-\mathcal{V}$ subsystem}
\label{AVsys}

The back reaction functions $\mathcal{A}$ and $\mathcal{V}$, or  equivalently $\mathfrak{A}$, satisfy a closed system of second order differential equations given by \eqref{Acal} and \eqref{Vcal}. This system can be solved by noting that the right hand side of \eqref{Acal} has no dependence on the back reaction functions\footnote{In general it is possible to write the right hand side of \eqref{Acal} in terms of functions lower order in a $Q_f$ than the ones appearing on the left hand side.}. Combining \eqref{Acal} and \eqref{Vcal} we find the solutions as

\vspace{2mm}
\begin{eqnarray}\label{solutionV}
\mathcal{V}'(r) &=&  Q^2_f V'_t(r) H(r) + \mathcal{O}\left(Q^4_f \right)\, ,\\
\label{solutionA}
\mathcal{A}(r) &=& \mathcal{A}(r_h) + \frac{Q^2_f}{x} \left[ L(r) + x \int^r_{r_h} dr' A'_t(r') H(r') \right]+ \mathcal{O}\left(Q^4_f \right)\ ,
\end{eqnarray}

\vspace{2mm}\noindent
where we used the boundary conditions \eqref{boundaryAdS} and introduced the following functions 
\vspace{2mm}
\begin{equation}
\begin{split}
H(r)&= \int^r_{\infty}  \frac{ dr' A_t(r') D(r')}{r^{'5} f(r') \sqrt{f(r') g(r')}}, \\ \\ L(r)&=\int^r_{r_h}  \frac{ dr' D(r')}{r^{'3} \sqrt{f(r') g(r')} Z_A(\phi(r'))},  \\ \\ D(r)&=\int^r_{r_h} dr' Z_0(\phi(r')) r' \sqrt{\frac{f(r')}{g(r')}}. 
\end{split}
\end{equation}
\noindent 
To evaluate $\mathcal{A}(r_h)$ we use the boundary condition 

\begin{equation}
\lim_{r \rightarrow \infty} \left[\frac{r}{r_h}\right]^{-\Delta} \mathcal{A} = 1,  \quad \quad \lim_{r \rightarrow \infty} \left[ \mathcal{A} - \frac{Q^2_f}{2x} \ln \left[ \frac{r}{r_h} \right] + \mathcal{O}\left(Q^4_f\right) \right] = 1. 
\end{equation}

\noindent and obtain

\vspace{2mm}
\begin{equation}
\mathcal{A}\left(r_h\right) = 1 + \frac{Q^2_f}{2x}  \lim_{r \rightarrow \infty} \left[  \ln \left( \frac{r}{r_h} \right) - 2L(r) - 2x \int^r_{r_h} dr' A'_t(r') H(r')  \right] + \mathcal{O}\left(Q^4_f \right).
\end{equation}
\noindent
Note that the limit is finite as the apparent log divergence is exactly cancelled by the divergent part of $L(r)$.

\section{Results}\label{conductivitiesSection}

\subsection{General results}

To find the linear response of the one point function it will be convenient to rewrite \eqref{vectorCurrentOne} in the reference frame of the solution at equilibrium:
\\
\begin{equation}
u= - dt + h_{ti} dx^i + \mathcal{O}\left(\partial\right), \quad \quad \tilde a = \tilde a_i dx^i,  + \mathcal{O}\left(\partial\right)\quad \quad \tilde v = \tilde v_i dx^i  + \mathcal{O}\left(\partial\right).
\end{equation}
\\
\noindent
We are interested in calculating the vector current 
\\
\begin{equation}\label{shearCurrent}
\begin{split}
\tilde J^i= 2 \kappa \tilde \epsilon^{\nu \mu \rho \sigma} \int^r_{r_h} dr \left[ A_t \mathcal{V}' - \mathcal{A} V'_t \right] \epsilon^{ijk} \partial_j \tilde a_k.
\end{split}
\end{equation}
\\
\noindent
Using \eqref{shearCurrent}, \eqref{16piG}, \eqref{solutionV}, and \eqref{solutionA} we find the linear response in the one point function of the vector current as 
\\
\begin{equation}\label{shearCurrent2}
\begin{split}
\langle J^i_V \rangle = \frac{-\kappa  \mu }{8 \pi G} \left[\mathcal{A}(r_h) +  \frac{Q^2_f}{x} \int^\infty_{r_h}dr V'_t \left( L(r) - xA_t H(r)  \vphantom{x \int^r_{r_h}dr' A'_t(r') H(r')}\right. \right. \\ \left. \left. + x \int^r_{r_h}dr' A'_t(r') H(r')       \right)     \right] B^{V,i} + \mathcal{O}\left(Q^4_f \right).
\end{split} 
\end{equation}
\\
\noindent

We now read off the anomalous conductivities from \eqref{shearCurrent2} as using 
\\
\begin{eqnarray}
\label{CME}
\sigma_{\text{\tiny CME}} &=& \sigma_{\text{\tiny CVE}}=0\, ,  \\
\label{CSE}
\sigma_{\text{\tiny CSE}}&=& 2 a_1 \left[\mu \mathcal{A}(r_h)   +  \frac{Q^2_f}{x} \int^\infty_{r_h}dr V'_t \left( L(r) - xA_t H(r) \vphantom{x \int^r_{r_h}dr' A'_t(r') H(r')} \right. \right. \nonumber\\ 
{}&&\left. \left. + x \int^r_{r_h}dr' A'_t(r') H(r')       \right)     \right] + \mathcal{O}\left(Q^4_f \right)\, , 
\end{eqnarray}
\\
\noindent
where we used \eqref{CScoefficients}. Equations \eqref{CME} and \eqref{CSE} constitute the main results of our paper. As a consistency check, we find that our expressions yield the known results (\ref{ConsCond}) in the limit $Q_f\rightarrow 0$. We  emphasize that to arrive at this result only universal expressions at the horizon were used. As such, these results are valid in a generic class of gravitational theories described by the generic action \eqref{holographicAction}. Finally, we note that the higher order corrections in $Q_f$ can  be computed by solving \eqref{Vcal} and \eqref{Acal} iteratively in a series expansion. The physical content of equation \eqref{CSE} depends on the choice of the background field, thus it will become more  clear when we present the examples in the next section. Nevertheless, we should note that the correction is of the expected universal form, namely proportional to $\mu$, and that it is present even in the absence of an axial gauge field background.

\section{Examples: chiral charge separation}\label{Examples}

In this section we provide three examples, two of which concern the chiral separation effect using our master formula (\ref{CSE}) and one calculating the chiral vortical separation effect---which we did not discuss in detail above---in a specific simple setting.  

\subsection{Reissner-Nordstrom Blackhole Background}
\label{RNBB}

\noindent The double charged Reissner-Nordstrom blackhole solution, with vanishing dilaton $\phi=0$ --- hence dilaton potential equals the cosmological constant $V=-12$ (with AdS lenght $\ell=1$) --- can easily be found by using the constants of motion \eqref{vectorConditions}, \eqref{axialCharges} and \eqref{gravitationalConstant} with $Q_f=0$:

\vspace{2mm}
\begin{equation}
f(r)=g(r) = 1 - \frac{M}{r^4} + \frac{\tilde Q^2}{r^6}= \frac{\left(r^2-r^2_h\right) \left( r^2 -r^2_+ \right) \left(r^2 - r^2_-\right) }{r^6},
\end{equation}

\begin{equation}
V_t = \mu \left[1 - \frac{r^2_h}{r^2} \right], \quad \quad A_t = \mu_5 \left[1 - \frac{r^2_h}{r^2} \right],
\end{equation}

\vspace{2mm}
\noindent where the mass $M$, the effective charge square $\tilde Q^2$, and the horizon radius $r_{\pm}$  are given by 

\vspace{2mm}
\begin{equation}
\tilde Q^2 = \frac{x r^4_h \left(\mu^2 + \mu^2_5 \right)}{3}, \quad \quad M= r_h^4 + \frac{\tilde Q^2}{r^2_h},
\end{equation}

\vspace{2mm}
\begin{equation}
r^2_{\pm}= -\frac{r^2_h \left[ 1  \pm  q \right]}{2} , \quad \quad q=\sqrt{1 + \frac{4 \tilde Q^2}{r^6_h} }.
\end{equation}

\vspace{2mm}
\noindent The temperature of this solution is 

\vspace{2mm}
\begin{equation}
T= \frac{2 r^2_h M - 3 \tilde Q^2 }{2 \pi r^5_h} = \frac{ 2r_h^6 - \tilde Q^2 }{2\pi r^5_h}.
\end{equation}

\vspace{2mm}
\noindent For this background the functions $D(r)$, $L(r)$ and $H(r)$ defined in section \ref{AVsys} read  

\vspace{2mm}
\begin{equation}
D(r)= \frac{r^2 - r^2_h}{2}, 
\end{equation}

\vspace{1mm}
\begin{equation}\label{aux1}
\begin{split}
L(r) = \frac{1}{4q} \left[ \left(\frac{r^2_-}{r^2_h} \right)\ln\left(\frac{ r^2 - r^2_-}{ r^2_h -r^2_- } \right) -\left(\frac{r^2_+}{r^2_h} \right) \ln\left(\frac{  r^2 - r^2_+}{r^2_h - r^2_+} \right)  \right],
\end{split}
\end{equation}

\vspace{1mm}
\begin{equation}\label{aux2}
\begin{split}
H(r)= \frac{\mu_5}{4 q^2 } \left[ \left(\frac{r^4_+}{r^4_h}\right) \frac{1}{r^2_+ - r^2} + \left(\frac{r^4_-}{r^4_h}\right) \frac{1}{r^2_- - r^2} + \frac{1-q^2}{2 q r^2_h} \ln \left[ \frac{r^2 - r^2_+}{r^2 - r^2_-} \right]  \right],
\end{split}
\end{equation}

\vspace{1mm}
\begin{equation}\label{aux3}
\begin{split}
\int^r_{r_h} A'_t H= \frac{\mu^2_5}{4 q^3 r^2_h} \left[  q \left(\frac{r^2_h}{r^2}- 1 \right) + \frac{q^2-3}{2} \ln\left(\frac{3-q}{3+q} \right) \right. \\ \left. \left(1 - \frac{r^2_h (q^2 -1)}{2 r^2} \right) \ln \left(\frac{r^2-r^2_-}{r^2-r^2_+} \right)\right].
\end{split}
\end{equation}

\vspace{2mm}

\noindent The functions $\mathcal{A}$ and $\mathcal{V}$ can be written down in terms of functions \eqref{aux1}-\eqref{aux3} and for the sake of clarity their relevant contribution to the CSE conductivity can be calculated separately

\vspace{2mm}
\begin{equation}
\begin{split}
\int^\infty_{r_h} \mathcal{A} V'_t = \mu + \frac{Q^2_f \mu}{x} \left\{  \frac{1}{8q}\left[ 3 \ln\left(\frac{3+q}{3-q} \right) + q \ln\left(\frac{9-q^2}{4} \right) \right]  \right.  \\ \left. + \frac{1}{8 q^3}\left( \frac{x \mu^2_5}{r^2_h} \right) \left[3q + \frac{9-q^2}{2} \ln\left(\frac{3-q}{3+q} \right)\right] \right\},
\end{split}
\end{equation}

\vspace{1mm}
\begin{equation}
\begin{split}
\int^\infty_{r_h} \mathcal{V'} A_t = -\frac{Q^2_f \mu}{x} \left(\frac{\mu^2_5 x}{r^2_h} \right) \frac{1}{8q^3} \left[3 q + \frac{9-q^2}{2} \ln \left(\frac{3-q}{3+q} \right) \right].
\end{split}
\end{equation}

\vspace{2mm}
We obtain the chiral separation conductivity as 

\vspace{2mm}
\begin{equation}\label{CSERN}
\begin{split}
\sigma_{\text{\tiny CSE}} = 2 a_1 \mu \left\{1 +  \frac{Q^2_f }{2x} \left(\frac{1}{4 q} \right) \left[ 3 \ln \left(\frac{3+q}{3-q} \right) + q\ln \left(\frac{9-q^2}{4} \right)  \right. \right. \\ \left. \left. + \frac{1}{q^2} \left(\frac{x \mu^2_5}{r^2_h} \right) \left[ 6q + (9-q^2) \ln \left(\frac{3+q}{3-q} \right) \right]  \right]  + \mathcal{O}\left(Q^3_f\right)  \right\}. 
\end{split}
\end{equation}

\vspace{2mm}
\noindent
We express \eqref{CSERN} in terms of physical parameters 
\begin{equation}
\label{physpar} 
\mu\equiv\frac{x \mu}{T}\, , \qquad \tilde \mu_5 \equiv \frac{ x \mu_5}{T}\, ,
\end{equation}
as 
\begin{equation}
\begin{split}
r_h=\frac{\pi T}{2} \left[1 + \sqrt{1+ \frac{2 \left( \tilde \mu^2 + \tilde \mu_5^2 \right)}{3 \pi^2}  } \right]  ,
\end{split}
\end{equation}

\begin{equation}
q= \sqrt{1 + \frac{16 x \left(\tilde \mu^2 + \tilde \mu^2_5 \right)}{ 3 \pi^2 \left(1 + \sqrt{1 + \frac{2 \left( \tilde \mu^2 + \tilde \mu_5^2 \right) }{3}} \right)^2  }   }.
\end{equation}

\vspace{2mm}
We observe that the CSE conductivity can be schematically rewritten as 
\\
\begin{equation}
\sigma_{\text{\tiny{CSE}}}= \sigma^{U}_{\text{\tiny{CSE}}} \left[  1 + \Delta\, \hat \sigma \left(\tilde \mu, \tilde \mu_5 \right)    + \mathcal{O}\left(Q^3_f \right) \right]\, ,
\end{equation}
\\
\noindent 
where $\sigma^{U}_{\text{\tiny{CSE}}}$ is the universal value of the chiral separation conductivity in the absence of dynamical gauge fields (\ref{ConsCond}) and we replaced the gluon-axial current coupling $Q_f$ by the anomalous dimension $\Delta$. 

\begin{figure}[H]
\centering
\begin{subfigure}{.5\textwidth}
  \centering
  \includegraphics[width=0.9\linewidth]{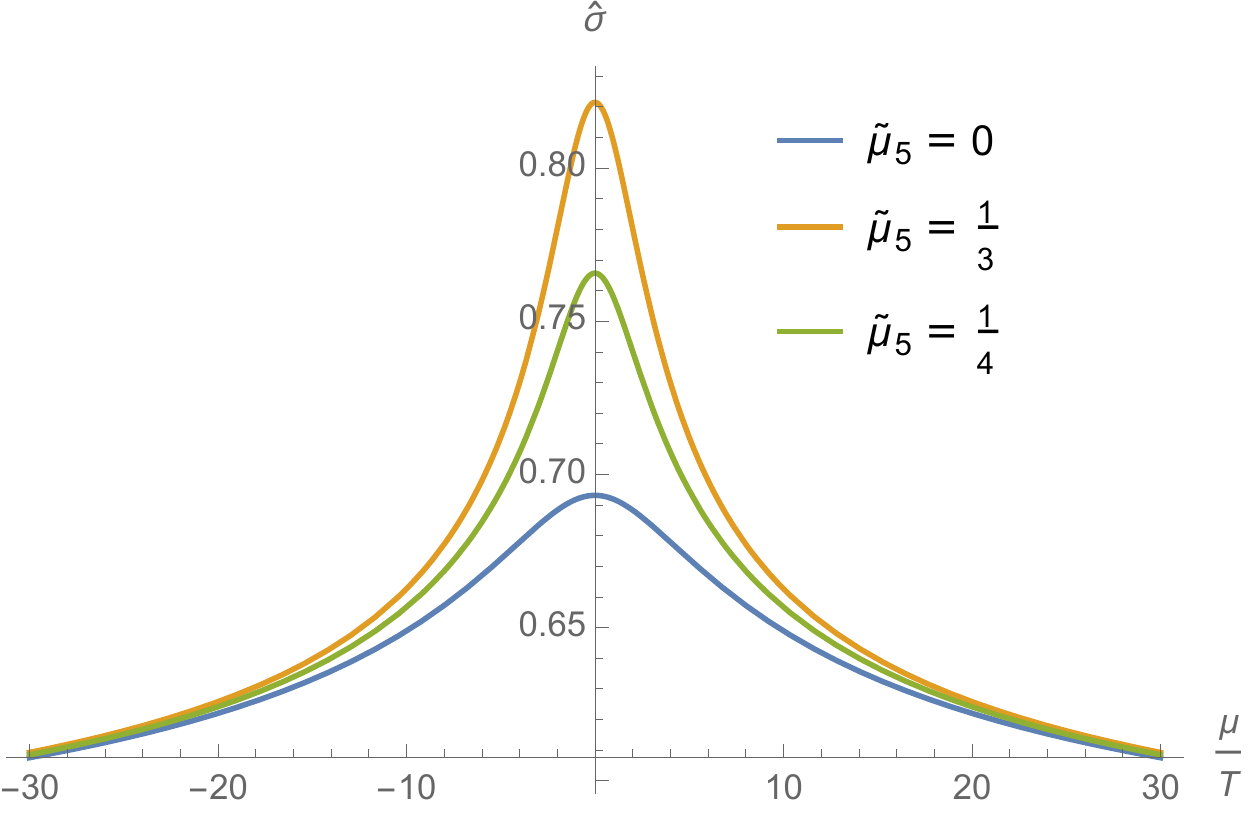}
\end{subfigure}%
\begin{subfigure}{.5\textwidth}
  \centering
  \includegraphics[width=0.9\linewidth]{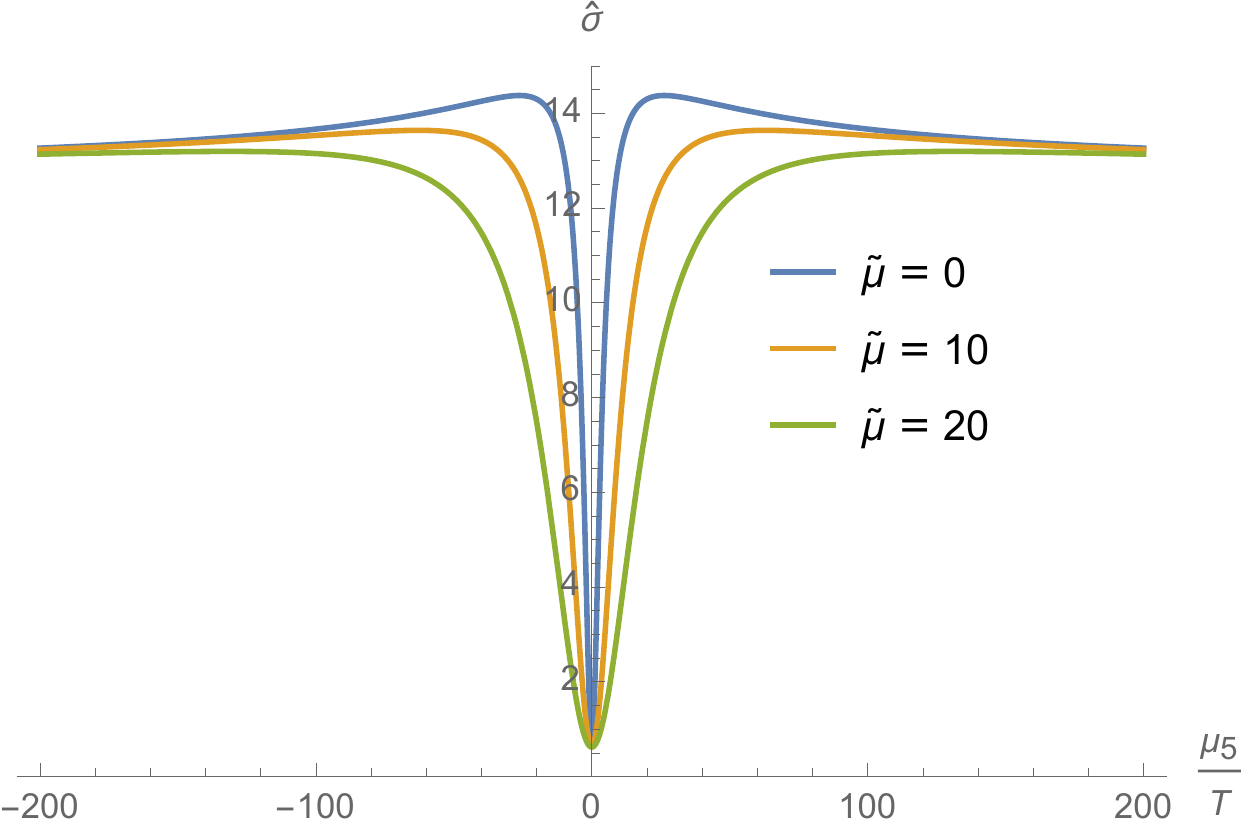}
\end{subfigure}
\caption{Left: Correction to the CSE conductivity due to dynamical gauge fields in the doubly charged ${\cal N}=4$ sYM plasma as a function of $\tilde \mu$ for: $\tilde \mu_5=0$ (blue), $\tilde \mu_5=\frac{1}{2}$ (red), $\tilde \mu_5 =2$. Right: Same as a function of $\tilde \mu_5$ for: $\tilde \mu=0$ (blue), $\tilde \mu=10$ (red), $\tilde \mu=20$ (green).}
\label{graphRN}
\end{figure}

We plot the correction $\hat \sigma$ in figure \ref{graphRN} as a function of $\tilde \mu$ for fixed  $\tilde \mu_5$ and as a function of $\tilde \mu_5$ for fixed $\tilde \mu$. 

\vspace{2mm}

\vspace{2mm}
\noindent 
We observe in figure \ref{graphRN} that  $\hat \sigma$ is symmetric, positive definite, and bounded from both above and below. It is also a decreasingly monotonic function of $\tilde \mu^2$, i.e.  for fixed $\tilde \mu_5$ it converges to a global minimum at $\tilde \mu \rightarrow \pm \infty$ and a global maximum at $\tilde \mu=0$. As a function of $\tilde \mu_5$ for fixed $\tilde \mu$ it attains a global maximum at some value $\tilde \pm \mu_{5c}$, a global minimum at $\mu_5=0$ and a local minimum at $\mu_5 \rightarrow \infty$. All in all we find that $\hat \sigma$ is bounded as 
\
\begin{equation}
\hat \sigma(\pm \infty,0)=\frac{\ln(3)}{2}  \leq \hat \sigma \left(\tilde \mu, \tilde \mu_5 \right) \leq 14.3765= \tilde \sigma(0,\pm 26.271 ).
\end{equation}

\vspace{2mm}
\noindent 
As the upper bound is substantial, in order to obey our assumption of a perturbative expansion in $\Delta$ we see that $\Delta$ may have to be very small. Otherwise the calculation is invalid for certain values of $\mu$ and $\mu_5$ that yield large $\hat\sigma$. 
For larger ${\cal O}(1)$ values of $\Delta$, a non-perturbative solution of the $\mathcal{A}-\mathcal{V}$ system of equations in section \ref{AVsys} will be needed, which can be obtained numerically for a given background.

 \subsection{Full analytic solution in the probe limit}
 \label{Ex2}

A non perturbative solution of the $\mathcal{A}-\mathcal{V}$ system can be obtained in the probe limit, that is, ignoring the metric fluctuations. This amounts to considering \eqref{Acal} as the only relevant equation in the system, i.e. setting $\mathcal{V}=0$ and disregarding Einstein's equations, namely ignoring equation \eqref{Vcal}. We consider the AdS-RN blackhole as the fixed background:

\begin{equation}\label{LandsteinerBackground}
\begin{split}
Z_A=Z_V=Z_0=1, \\
f(r)=g(r)= 1 - \frac{r^4_h}{r^4}. \\
V_t = \mu \left[ 1 - \frac{rh^2}{r^2} \right]
\end{split}
\end{equation}

\noindent In this limit the CSE conductivity becomes

\begin{equation}
\sigma_{\text{\tiny{CSE}}}=2 a_1 \int^\infty_{1}  dr \mathcal{A} V'_t,
\end{equation}

\noindent where the position of the horizon can be set to $r_h=1$. $\mathcal{A}$ satisfies

\begin{equation}
\partial_r \left[ \sqrt{-\mathcal{G}} Z_A g \mathcal{A}' \right] = \frac{Q^2_f}{x} \frac{\sqrt{-\mathcal{G}} Z_0 \mathcal{A}}{r^2}.
\end{equation}

\noindent
Regularity at the horizon together with the asymptotic behavior $\lim_{r \rightarrow \infty} \mathcal{A}=\left(\frac{r}{r_h}\right)^{\Delta}$ fixes the solution as

\begin{equation}
\mathcal{A}= D_1(\Delta) \left[ F_1(\Delta,r) + D(\Delta) r^2 F_2(\Delta,r) \right],
\end{equation}

\vspace{1mm}
\noindent where functions $F_1$ and $F_2$ are defined as

\begin{equation}
F_1\equiv  {}_2F_1 \left(-\frac{\Delta}{4}, \frac{2 +\Delta}{4}, \frac{1}{2}, r^4 \right) \quad \quad F_2 \equiv  {}_2F_1\left( - \frac{\Delta-2}{4}, \frac{4+\Delta}{4},\frac{3}{2}, r^4\right), 
\end{equation}

\vspace{2mm}
\noindent  and the functions $D(\Delta)$ and $D_1(\Delta)$ read 

\vspace{2mm}
\begin{eqnarray}
D_1(\Delta) &=& \frac{(-1)^{-\Delta/4}}{\sqrt{\pi} \Gamma\left(\frac{\Delta+1}{2} \right) } \left[\frac{1}{\left[\Gamma \left(\frac{\Delta+2}{4}\right) \right]^2} - \frac{i D(\Delta)}{2\left[\Gamma\left(\frac{\Delta+4}{4}\right) \right]^2}  \right]^{-1} \\
D(\Delta) &=& \frac{\Delta}{2} \frac{\Gamma\left(\frac{2-\Delta}{4} \right) \Gamma\left(\frac{4+\Delta}{4} \right)}{\Gamma \left(\frac{2+\Delta}{4} \right) \Gamma \left(\frac{4-D}{4} \right)}
\end{eqnarray}

\vspace{2mm}
The CSE conductivity is then given by 

\vspace{2mm}
\begin{equation}\label{CSELandsteiner}
\sigma_{\text{\tiny{CSE}}}= 2 a_1 \mu D_1 \left(\Delta\right)\int_{1}^\infty \left[ \frac{F_1(\Delta,r)}{r^3} + D(\Delta) \frac{F_2(\Delta,r)}{r} \right] 
\end{equation}

\begin{figure}[h]
\centering
  \includegraphics[width=0.7\linewidth]{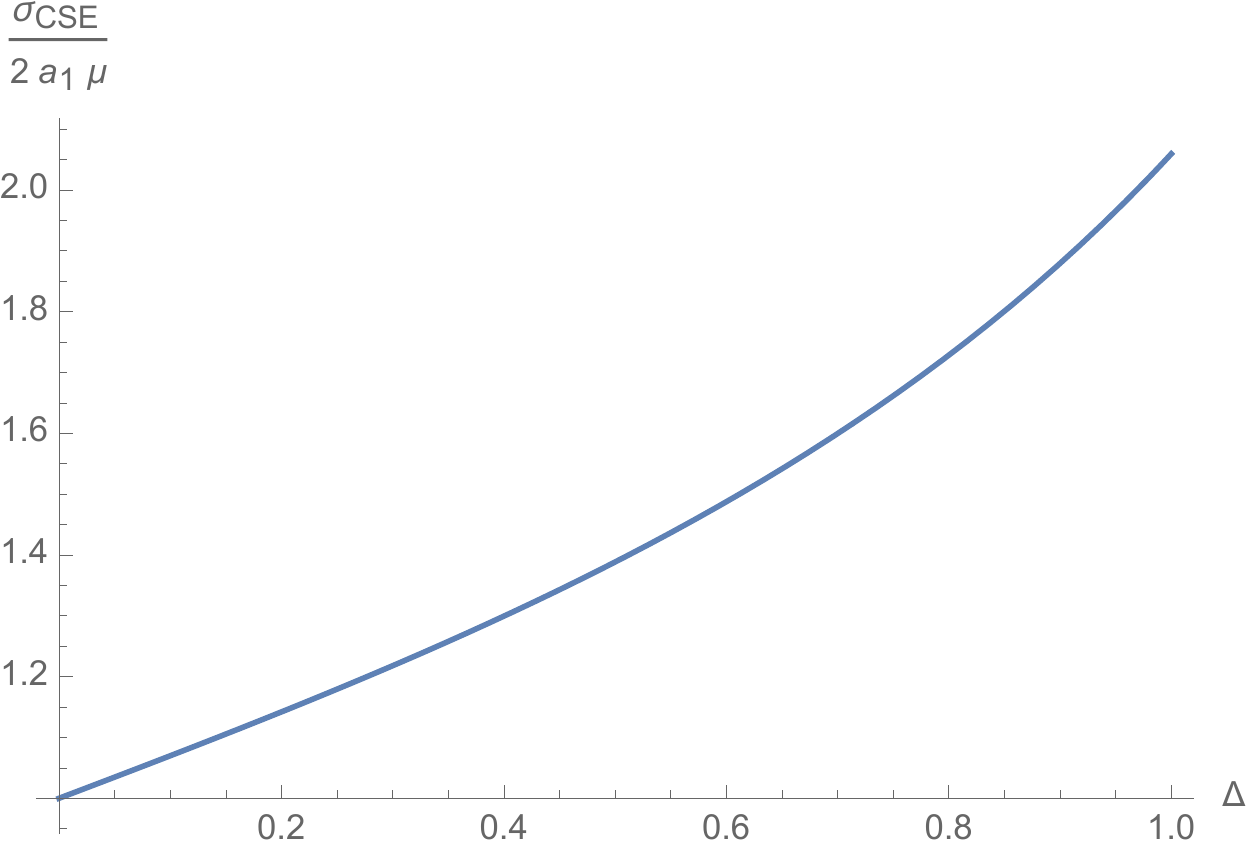}
  \caption{Plot of the correction to the CSE conductivity as a function of the anomalous dimension $\Delta$ in a fixed AdS blackhole background.}
\label{figureLandsteiner}
\end{figure}

\vspace{2mm}
\noindent The integrals in \eqref{CSELandsteiner} can be done analytically and the final result for the conductivity can be expressed in terms of hypergeometric and Meijer-G functions. Nevertheless it is more informative to plot it as a function of $\Delta$. This is shown in figure \ref{figureLandsteiner}.

The result is consistent with the numerical calculation of the CSE conductivity done in \cite{LandsteinerStuckleberg}. It should be noted that the solution starts to deviate from the linear approximation $\sigma_{\text{\tiny{CSE}}}= 2 a_1 \mu \left[1 + \Delta \ln(2) \right]$ at around $\Delta \sim 0.35$.

\subsection{Chiral vortical separation effect in neutral conformal plasma}

Calculation of the CVSE conductivity, that is the chiral separation effect due to vortices, is generically harder as it requires solving the axial perturbation equations fully. There is a specific example where this can be done analytically at first order in $Q_f$, that is for $\mu=\mu_5=0$ in the background studied in \eqref{LandsteinerBackground}. Considering $\tilde v=\tilde a=0$, the $\mathcal{O}\left(Q^2_f \right)$ correction to the CVSE conductivity can be calculated. In this limit the relevant $\mathcal{O}\left(\partial \right)$ equation becomes

\begin{equation}\label{CVSEequation}
\partial_r \left( \sqrt{-\mathcal{G}} Z_A g  \alpha'_{\omega} \right) = - \lambda \partial_r \left(\frac{r^4  g f'^2}{2f } \right) + \frac{Q^2_f}{x} \frac{\sqrt{-\mathcal{G}} Z_0 \alpha_\omega}{r^2} 
\end{equation}

\noindent We define a covariant bulk axial current $\tilde J^\nu_A$ as 

\begin{equation}
\tilde J^\nu_A = \sqrt{-\mathcal{G}} x Z_A F^{A,r \nu} = \sqrt{-\mathcal{G}} x Z_A g  \alpha'_\omega \omega^\nu 
\end{equation}

\noindent from which we obtain the  renormalized axial one point function \cite{LandsteinerStuckleberg} 

\begin{equation}\label{renormalizedAxial}
16 \pi G_N \langle J_A^\nu \rangle = \lim_{r \rightarrow \infty} \left[- \tilde J^\nu_A  + r \Delta \alpha \omega^\nu  \right] 
\end{equation}

\noindent The solution to \eqref{CVSEequation} in a $Q_f$ expansion is given by 

\begin{equation}\label{solutionAlpha}
\alpha_\omega = \alpha_1(r) + \alpha_2(r) Q^2_f + \mathcal{O}\left(Q^3_f \right),
\end{equation}

\noindent with 

\begin{equation}
\begin{split}
\alpha_1 &= - \frac{\lambda 8\pi T}{r_h} \left[ \frac{r^4_h }{4 r^4} + \frac{1}{2}\ln \left(1 + \frac{r^2_h}{r^2} \right) \right] \\
\alpha_2 &= (16 \pi T r_h ) \lambda \left[ \frac{1}{8} \left(1-\frac{r^2_h}{r^2} \right) + \frac{r^2}{4 r^2_h} \ln \left(\frac{r^2_h}{r^2} +1 \right) + \frac{1}{2} \ln \left(\frac{r}{r_h} \right) \right. \\ {}&\left.  + \frac{1}{4} \ln \left(1 + \frac{r^2_h}{r^2} \right) - \frac{\ln 2}{2} \right]
\end{split}
\end{equation}
 
\noindent Using the solution \eqref{solutionAlpha} together with \eqref{renormalizedAxial} we finally obtain the CVSE conductivity as 

\begin{equation}\label{CVSEconductivity}
\sigma_{\text{\tiny{CVSE}}}= 8 \pi^2 a_4 T^2 \left[1 + \frac{\Delta(\Delta+2)}{2 } \left(\ln 2 - \frac{1}{4} \right) + \mathcal{O}\left(Q^3_f \right) \right] 
\end{equation}

\noindent We observe that the $T^2$ contribution to the CVSE conductivity also receives correction from dynamical gluons in strongly coupled ${\cal N}=4$ sYM.  Just like the CSE, this correction is positive.

\section{Discussion} \label{Conclusions}

We developed a semi-analytic procedure to calculate the chiral magnetic, chiral separation, and chiral vortical conductivities in the presence of dynamical gauge fields in strongly interacting chiral gauge theories in the Veneziano large color, large flavor limit using the holographic correspondence. We find that, while the CME and CVE conductivities do not receive corrections, the CSE conductivity does, and it is given by two background functions $\{\mathcal{A},\mathcal{V}\}$ that satisfy a closed and coupled system of differential equations. We solved this system of equations in a perturbative expansion in the anomalous dimension of the axial current $\Delta$ and obtained an analytic expression at first order.

Quite generally --- for theories that can be described by two-derivative gravity --- we find that the correction to the CSE conductivity due to the dynamical gauge fields are positive definite. This is to be contrasted with negative corrections  obtained in some lattice calculations \cite{Yamamoto:2011gk, Braguta:2013loa}. There is no clash in these results however, since our calculation is strictly valid in the Veneziano limit, at large 't Hooft coupling $\lambda$ and in a class of gauge theories that are related to but not the same as QCD. It will be interesting to extend our results beyond these limitations, in particular to determine the $1/\lambda$ corrections. Whereas existing holographic studies \cite{GursoyTarrio, Grozdanov} indicate no deviation from the universal values, i.e. no $1/\lambda$ corrections in the absence of dynamical gauge fields, there is no analogous result when such dynamical gauge fields are taken into account. This is an open problem. 

The example we provide in section \ref{RNBB} show that the validity of the small $\Delta$ expansion should be checked carefully. There is a range of parameters $\mu$ and $\mu_5$ where the perturbative expansion makes sense only for very small choices of $\Delta$. This prompts us to look for alternatives to the small $\Delta$, or equivalently small $Q_f$ expansion. One such powerful approach would solving the $\mathcal{A}-\mathcal{V}$ system of differential equations. This provides a straightforward way to obtain, at least numerically, the non perturbative solutions. We provided an example of such a calculation in the case of AdS blackhole with no backreaction in section \ref{Ex2}. This particular example was first discussed in \cite{LandsteinerStuckleberg} our  analytic results are consistent with the numerical results in this paper. We should stress that the method we use is very different than the method of \cite{LandsteinerStuckleberg}. Whereas \cite{LandsteinerStuckleberg} solves the second order fluctuation equations numerically and obtain the conductivity from the Kubo formula, we solve the second order $\mathcal{A}-\mathcal{V}$ ODE system, which seems to  provide a simpler way to obtain the conductivity with full back reaction either numerically or analytically. 

Finally, we should remark that we treated the anomalous dimension of the axial current $\Delta$ as a (small) tuneable parameter in our model. In reality, $\Delta$ itself should be determined in terms of the anomaly coefficients \cite{Adler:2004qt}. Whereas, this can be achieved in perturbative QFT, it is not at all clear how to proceed at strong coupling. We suspect however a holographic relation between $\Delta$ and the anomaly coefficients might exist. We plan to return this problem in the future.

\begin{acknowledgements}
We thank Aron Jansen, Karl Landsteiner, David Mateos and Fran Pena-Benitez for useful discussions and especially Javier Tarrio who has participated at the early stages. UG's work is supported in part by the Netherlands Organisation for Scientific Research (NWO) under VIDI grant 680-47-518, the Delta Institute for Theoretical Physics (D-ITP) funded by the Dutch Ministry of Education, Culture and Science (OCW), the Scientific and Technological Research Council of Turkey (TUBITAK). UG is grateful for the hospitality of the Bo\~gazi\c ci University and the Mimar Sinan University in Istanbul. DG is supported in part by CONACyT through the program Fomento, Desarrollo y Vinculacion de Recursos Humanos de Alto Nivel.
\end{acknowledgements}

\appendix
\section{Geometrical Data at $\mathcal{O}\left(\partial^0 \right)$}\label{appendixA}

The Christoffel symbols for an Ansatz of the form $ds^2=\mathcal{G}_{rr}(r) dr^2 + \mathcal{G}_{\mu \nu}(r,x) dx^\mu dx^\nu $ are schematically given by

\begin{equation}
\begin{split}
\Gamma^r_{rr}&= \frac{1}{2}G^{rr} G'_{rr}+ \mathcal{O}\left(\partial \right) \quad \quad 
\Gamma^r_{r\nu} = \mathcal{O}\left(\partial\right) \quad \quad 
\Gamma^r_{\mu \nu}=-\frac{1}{2} G^{rr} G'_{\mu \nu} + \mathcal{O}\left(\partial \right)  \\
\Gamma^\mu_{rr}&=\mathcal{O}\left(\partial\right) \quad \quad
\Gamma^\mu_{r\nu}= \frac{1}{2} G^{\mu \rho} G'_{\rho \nu} + \mathcal{O}\left(\partial \right) \quad \quad
\Gamma^\mu_{\rho \sigma}= \mathcal{O}\left(\partial  \right)
\end{split}
\end{equation}
It follows that only $\{\Gamma^r_{rr},\Gamma^r_{\mu \nu}, \Gamma^\mu_{r \nu} \}$ contribute at the zeroth order. Therefore the only independent non-vanishing components of the Riemann tensor at this order are $\{R^{r}_{\hphantom{r}\mu r \nu}, R^\mu_{\hphantom{\mu} \nu  \rho \sigma} \} $. Then the following schematic expressions for the components of the Ricci tensor are obtained, 
\vspace{2mm}
\begin{equation}
\begin{split}
R_{rr}&=  - \left(\Gamma^\rho_{r \rho} \right)' + \Gamma^r_{rr} \Gamma^\rho_{\rho r} - \Gamma^\sigma_{r \rho} \Gamma^\rho_{r \sigma} + \mathcal{O}\left( \partial \right) \\
R_{r\mu}&= \mathcal{O}\left( \partial \right)\\
R_{\mu \nu} &= \left(\Gamma^r_{\mu \nu} \right)' + \Gamma^r_{\mu \nu} \left(\Gamma^r_{rr} + \Gamma^\rho_{\rho r} \right) - 2 \Gamma^r_{\rho (\mu} \Gamma^\rho_{\nu)r} + \mathcal{O}\left( \partial \right)
\end{split}
\end{equation}
The particular ansatz \eqref{metricAnsatz} leads to the following the non-vanishing Christofell symbols 
\begin{equation}
\begin{split}
\Gamma^r_{rr}&= -\left( \ln\left[r \sqrt{g} \right]\right)'\\
\Gamma^r_{\mu \nu} &= \frac{r^2 g}{2} \left[\left(r^2 f \right)' u_\mu u_\nu - \left(r^2 \right)' \Delta_{\mu \nu} - \left(r^2 \mathfrak{A} \right)' u_{(\mu}\tilde a_{\nu)} - \left(r^2 \mathscr{A} \right)' \tilde a_\mu \tilde a_\nu \right]\\
\Gamma^\mu_{r \nu} &=  - \left(\ln\left[r \sqrt{f} \right] \right)' u^\mu u_\nu  + \left(\ln[r] \right)' \Delta^\mu_\nu   + \left(\frac{\mathscr{A'}}{2}\right) \tilde a^\mu \tilde a_\nu \\  {}&+ \frac{1}{2}\left( \frac{\left(r^2 \mathfrak{A} \right)'}{fr^2} - \frac{(r^2)'\mathfrak{A}}{fr^2} \right) u^\mu \tilde a_\nu + \frac{1}{2}\left(\frac{\left(r^2 \mathfrak{A} \right)'}{r^2} - \frac{\left(fr^2\right)' \mathfrak{A}}{fr^2} \right) \tilde a^\nu u_\mu +  \mathcal{O}\left(Q^4_f \right)\, ,
\end{split}
\end{equation}
which lead to the following useful identities
\begin{equation}
\begin{split}
\Gamma^\rho_{r \rho} & = \left( \ln \left[r^{D-1} \sqrt{f} \right]  + \frac{\mathscr{A} \tilde a^2}{2}\right)' + \mathcal{O}\left(Q^4_f \right)\\
\Gamma^\sigma_{r \rho} \Gamma^\rho_{r \sigma} &= \left[\left(\ln\left[r \sqrt{f} \right] \right)' \right]^2 + \frac{D-2}{r^2} + \frac{\mathscr{A}'\tilde a^2}{r} + \mathcal{O}\left(Q^4_f \right)\\
\Gamma^r_{\rho (\mu} \Gamma^\rho_{\nu) r} &= \frac{r^2g}{2} \left\{ (r^2 f)' \left( \ln\left[r \sqrt{f} \right] \right)' u_\mu u_\nu - \left(r^2 \right)' \left(\ln[r] \right)' \Delta_{\mu \nu}   \right. 
\\ {} &\left. \left[ \frac{(r^2 f)' (r^2)' \mathfrak{A}}{2r^2 f} - \left(r^2 \mathfrak{A} \right)' \left( \ln \left[ r^2 \sqrt{f} \right] \right)' \right] \tilde a_{(\mu} u_{\nu)}  
- \left[2 r \mathscr{A} \right]' a_\mu \tilde a_\nu\right\} + \mathcal{O}\left(Q^4_f \right)\, .
\end{split}
\end{equation}
with $D$ being the dimension of the bulk geometry.
Using these identities the only non-vanishing Ricci tensor components are found as
\begin{equation}\label{RicciRR}
\begin{split}
R_{rr} &=- \left( \ln\left[r^{D-1} \sqrt{f} \right]  + \frac{\mathscr{A} \tilde a^2}{2} \right)'' - \left( \ln\left[r^{D-1} \sqrt{f} \right]  + \frac{\mathscr{A} \tilde a^2}{2} \right)'\left( \ln\left[r \sqrt{g} \right]\right)' \\
{}&- \left[\left(\ln\left[r \sqrt{f} \right] \right)' \right]^2 - \frac{D-2}{r^2} - \frac{\mathscr{A}' \tilde a^2}{r} + \mathcal{O}\left(Q^4_f \right)\, .
\end{split}
\end{equation}

\vspace{4mm}
\begin{equation}\label{RicciMM}
\begin{split}
R_{\mu \nu} &= \left\{ \left[ \frac{r^2 g}{2} \left(r^2 f \right)' \right]'  + \left[ \ln\left(  \frac{r^{D-4}}{ \sqrt{fg} } \right)  + \frac{\mathscr{A}\tilde a^2}{2}  \right]' \left[  \frac{r^2 g}{2} \left(r^2 f \right)' \right]      \right\} u_\mu u_\nu \\
{}&- \left\{  \left[\frac{r^2 g}{2} \left(r^2 \right)'\right]'  + \left[ \ln\left( r^{D-4} \sqrt{\frac{f}{g}} \right) + \frac{\mathscr{A} \tilde a^2}{2}  \right]' \left[\frac{r^2 g}{2} \left(r^2 \right)' \right]   \right\} \Delta_{\mu \nu} \\
{}&- \left\{ \left[ \frac{r^2 g}{2} \left(r^2 \mathfrak{A} \right)' \right]' + \left[ \ln \left(\frac{r^{D-6}}{\sqrt{fg}} \right) \right]' \left[ \frac{r^2 g}{2} \left(r^2 \mathfrak{A} \right)'  \right]  + \frac{(r^2f)' (r^2)'g \mathfrak{A}}{2f}  \right\} u_{(\mu} \tilde a_{\nu)} \\
{}&- \left\{ \left[ \frac{r^2 g}{2} \left(r^2 \mathscr{A} \right)' \right]'  + \left[ \ln \left( r^{D-1}\sqrt{\frac{f}{g}} \right) \right]'\left[ \frac{r^2 g}{2} \left(r^2 \mathscr{A} \right)'\right] - 2 r^2 g \left[r \mathscr{A} \right]' \right\}  \tilde a_{\mu} \tilde a_{\nu}\, .
\end{split}
\end{equation}

\section{Gravitational Conserved Charge}\label{appendixB}

At the zeroth order the radial and spatial parts of Einstein's equations do no mix, giving rise to four sets of independent equations. As detailed in appendix \ref{appendixA}, the radial projection yields

\vspace{2mm}
\begin{equation}
R_{rr}= \frac{\phi'^2}{2} + \frac{\rho}{2 r^2 g} - \frac{x}{2f} \left[Z_V V^{'2}_t + Z_A A^{'2}_t \right] + \mathcal{O}\left(\partial \right),
\end{equation}
\noindent
where $R_{rr}$ is shown in \eqref{RicciRR} and $\rho$ is defined as  

\begin{equation}
\rho \equiv \left[\frac{2V}{3} + \frac{xg \left(Z_V V^{'2}_t + Z_A A^{'2}_t \right)}{3f} \right].
\end{equation}
\noindent
\vspace{2mm}
On the other hand, the projection along $\Delta^{\mu \nu}$ reads

\vspace{2mm}
\begin{equation}\label{tangentPro}
\begin{split}
-3 \left[ \frac{1}{r}\sqrt{\frac{g}{f}} \left[ r^4 \sqrt{fg} \right]' + \frac{\rho r^2}{2} \right] &= \tilde a^2 \left\{ \frac{3 \left[ r^3 g \mathscr{A}'  + \rho \mathscr{A} \right]}{2}  \vphantom{\frac{Q^2_f Z_0}{2}} \right. \\ {}&\left.+ \left[ \frac{r^2 g}{2} \left(r^2 \mathscr{A} \right)' \right]'   + \left[ \ln \left( r^4 \sqrt{\frac{f}{g}} \right) \right]'\left[ \frac{r^2 g}{2} \left(r^2 \mathscr{A} \right)'\right] \right. \\ {}&\left.  - 2 r^2 g \left[r \mathscr{A} \right]' + \frac{Q^2_f Z_0}{2} \right\} + \mathcal{O}\left(\partial,Q^4_f\right),
\end{split}
\end{equation}
\noindent
From this equation it follows that 

\begin{equation}
\rho= -\frac{2}{r^{3}} \sqrt{\frac{g}{f}} \left[ r^{4} \sqrt{fg} \right]' + \tilde \rho(\tilde a^2 Q^2_f, \partial),
\end{equation}

\vspace{2mm}
\noindent 
where $\tilde \rho$ is some function. Similarly, the projection along $u^\mu u^\nu$ is given by 

\vspace{2mm}
\begin{equation}\label{velocityPro}
\begin{split}
\frac{\sqrt{fg}}{r} \left[ \frac{r^{3} (r^2 f)'}{2} \sqrt{\frac{g}{f}} \right]' + \frac{\mathscr{A}'}{2} \left[\frac{r^2 g}{2} \left(r^2 f \right)' \right]\tilde a^2 &= \frac{Q^2_f Z_0}{2} A^2_t - \frac{r^2 f \rho}{2}\\ &+ \frac{x Z_V V^{'2}_t}{2} + \frac{x Z_A A^{'2}_t}{2} + \mathcal{O}\left(Q^4_f, \partial \right).
\end{split}
\end{equation}

\vspace{2mm}
\noindent
Equations \eqref{tangentPro} and \eqref{velocityPro} can be combined as 

\vspace{2mm}

\begin{equation}\label{gravitationalConstantAppendix}
\left[ r^5 f' \sqrt{\frac{g}{f}} - x \left(A_t Q_5 + Q V_t \right) \right]' = \hat \rho \left( \tilde a^2 Q^2_f, \partial \right),
\end{equation}

\vspace{2mm}
\noindent where $\hat \rho$ is some function.
\bibliographystyle{ieeetr} 
\bibliography{gluonAnomalyD4}

\end{document}